\documentclass[doublespacing,onecolumn,extra]{gji}

\usepackage{timet}
\usepackage{xcolor}
\usepackage{amssymb}
\usepackage{amsmath}
\usepackage{amsfonts}
\usepackage{mathrsfs}
\usepackage{bm}
\usepackage{caption}
\usepackage{subcaption}
\usepackage{graphicx}
\usepackage{tikz}
\usepackage{enumitem}
\usepackage{listings}
\usepackage[final]{changes}
\usepackage[percent]{overpic}

\usepackage{algorithm2e}
\usepackage[colorlinks=true, linkcolor=blue, citecolor=blue, urlcolor=blue]{hyperref}
\usepackage{cleveref}
\usetikzlibrary{arrows.meta,calc,positioning}
\tikzset{
  ann/.style={-Latex, line width=0.7pt},
  lbl/.style={anchor=north west, inner sep=1pt, font=\normalfont\small}
}
\newcommand{\dif}{\,\mathrm{d}}
\newcommand{\mb}{\mathbf}
\newcommand{\R}{\mathbb{R}}

\title[FE methods for planetary gravitation]
{Efficient parallel finite-element methods for planetary gravitation: DtN and multipole expansions}

\author[Z. Yu \emph{et al.}]
{Ziheng Yu$^1$\thanks{zy296@cam.ac.uk}, Alex D.C. Myhill$^1$,  and David Al-Attar$^1$ \\
$^1$ Bullard Laboratories, Department of Earth Sciences, University of Cambridge, UK CB3 0EZ
}
\date{Received ; in original form }
\volume{}
\pubyear{{\the\year{}}}

\begin{document}
\label{firstpage}
\maketitle

\begin{summary}
The Poisson equation governing a planet's gravitational field is posed on the unbounded domain, $\R^3$, whereas finite-element computations require bounded meshes. We implement  and compare three strategies for handling the infinite exterior in the finite-element method: (i) naive domain truncation; (ii)  Dirichlet-to-Neumann (DtN) map on a truncated boundary; (iii) multipole expansion on a truncated boundary. While all these methods are known within the geophysical literature, we discuss their parallel implementations within modern open-source finite-element codes, focusing specifically on the widely-used MFEM package. 
We consider both calculating the gravitational potential for a static density structure and computing the linearised perturbation to the potential caused by a displacement field—a necessary step for coupling self-gravitation into planetary dynamics. In contrast to some earlier studies, we find that the domain truncation method can provide accurate solutions at an acceptable cost, with suitable coarsening of the mesh within the exterior domain. Nevertheless, the DtN and multipole methods provide superior accuracy at a lower cost within large-scale parallel geophysical simulations despite their need for non-local communication associated with spherical harmonic expansions. The DtN method, in particular, admits an efficient parallel implementation based on an MPI-communicator limited to processors that contain part of the mesh's outer boundary. A series of further illustrative calculations are provided to show the potential of the DtN and multipole methods within realistic geophysical modelling.
\end{summary}


\section{Introduction}\label{sec:intro}

The gravitational potential generated by mass distributions satisfies Poisson's equation on the unbounded domain $\mathbb{R}^3$, with growth conditions at infinity rendering the potential unique up to the addition of a constant \citep[e.g.][]{kellogg1953foundations, dahlen1999theoretical}. This problem occurs frequently in geophysical and planetary science applications, including glacial isostatic adjustment (GIA) modelling which constitutes the principal motivation for this study. For the GIA problem, it is 
necessary to solve for the static gravitational field of the equilibrium planet, and to account for linearised changes to the gravitational field associated with load-induced deformation via coupling of Poisson's equation to the equations for quasi-static viscoelasticity. Within spherically symmetric planets, spectral methods can be applied for GIA modelling which allow for a simple and exact treatment of the exterior part of Poisson's equation \citep[e.g.][]{wu1982viscous,tanaka2011spectral,al2014sensitivity}. Within recent years, however, interest within the GIA community has shifted towards laterally heterogeneous earth models using finite-element or finite-volume methods \citep[e.g.][]{wu2004using,latychev2005glacial, metivier2006mantle,wahr2013computations,huang2023commercial}. In this context the exterior part of Poisson's equation presents a more significant challenge, and it has typically been addressed in an approximate manner or using inefficient numerical approaches that contribute significantly to the overall calculation time. 

The difficulty in solving Poisson's equation for the gravitational potential using finite-element (or similar) methods is that computations must be restricted to bounded meshes. This necessitates the introduction of modified  boundary conditions that reproduce the correct exterior physics to a sufficient level of accuracy. Elliptic equations, such as the Poisson problem, are especially sensitive to the manner in which the exterior domain is handled, because their solutions decay rather slowly with distance from the interior domain \citep[e.g.][]{gilbarg2001}. This challenge links directly to a long history of work on non-reflecting boundary conditions in applied mathematics and computational mechanics \citep[e.g.][]{engquist1977absorbing, bayliss1980radiation, higdon1986absorbing, givoli1991non}.

A variety of strategies have been developed to represent the infinite exterior domain within finite-element solutions of Poisson's equations performed on a bounded mesh. The most straightforward is naive truncation of the computational domain to a large but finite size, with  either homogeneous Dirichlet or Neumann conditions imposed on the potential at the outer boundary. The error associated with this approximations is known to decrease rather slowly with the size of the computational domain, and hence this simple approach has sometimes been discounted \citep[e.g.][]{gharti2017spectral}. The computational cost associated with increasing the size of the  domain is, however, controlled most directly by the number of degrees of freedom within the finite-element discretisation, and this number need not grow dramatically with  the size of the domain if a suitable mesh coarsening scheme is employed. 

A more sophisticated approach is the use of the Dirichlet-to-Neumann (DtN) map, which eliminates the exterior domain by providing the relation between the potential and its normal derivative on the boundary of the computational domain \citep[e.g.][]{keller1989exact,givoli1989finite,chaljub2004spectral,metivier2006mantle}. On spherical boundaries this map is diagonal in a spherical harmonic basis, making it particularly attractive for geophysical problems \citep[e.g.][]{chaljub2004spectral}. The accuracy of this method is determined principally by the truncation order chosen within the spherical harmonic expansion, while the need to link finite-element and spherical harmonic representations of the boundary potential presents a non-trivial complication to its efficient parallel implementation. The nature of the Dirichlet-to-Neumann term within the weak formulation of Poisson's equation also means that their implementation within open source finite-element packages is not immediate. 
 
A related but distinct idea is the multipole expansion method, which originates from classic potential theory, and is based upon the representation of the exterior potential as a sum over multipole moments generated by the density distribution \citep[e.g.][]{bayliss1982boundary, may2011optimal,van2021modelling}. The coefficients of this expansion are computed through a set of volume integrals over the planet's interior, and the truncated series is used to determine non-homogeneous boundary conditions for the finite-element solution on an artificial spherical boundary within the exterior domain. As with DtN, the accuracy of this method is governed by the order used within the multipole expansion, while again there are challenges in linking finite-element and spherical harmonic representations of the potential and with its implementation within open-source FEM packages.

An alternative philosophy is the infinite element method (IEM), in which the exterior domain is collapsed into a thin computational layer adjoining the source region \citep[e.g.][]{beer1981infinite,zienkiewicz1983novel,bettess1991infinite}. There are two variants of this approach. The first is known as `co-ordinate ascent' and involves a singular mapping from a finite layer onto the exterior domain. The result of this process is a modified Poisson problem posed
on a bounded domain but with spatially variable coefficients that are singular on the external boundary. The numerical solution of the modified problem proceeds using standard finite-element basis functions, but with a modified quadrature rule needed to handle the integrable singularity on the outer boundary. Alternatively, within the `co-ordinate descent' technique, the form 
of the exterior Poisson equation is retained, but non-standard basis functions are used in this domain that decay in a prescribed manner from the interior body (typically with inverse powers of radius).
The contribution of the exterior terms to the overall linear system can be evaluated by again using a singular mapping of a finite layer onto the exterior domain coupled to a non-standard quadrature 
scheme. A recent variant of this method is the spectral infinite element method \citep[e.g.][]{gharti2017spectral,gharti2018spectral,gharti2019spectral, gharti2023spectral}, 
this being a form of co-ordinate descent which is designed to couple naturally to spectral element discretisations that are widely used in computational seismology \citep[e.g.][]{komatitsch2002spectral}. 
A clear advantage of the infinite element approach is that it retains the local nature of finite-element discretisation, with this property aiding significantly within efficient parallelisation. A disadvantage, however, is that the accuracy of the method is tied to the decay rates chosen either explicitly (in co-ordinate descent) or implicitly (in co-ordinate ascent) for the basis functions in the exterior domain, with a chosen set of basis functions not necessarily being sufficient to ensure a desired level of accuracy. For example, if the potential on the surface 
of the interior domain is characterised by spherical harmonic powers up to degree $\ell$, then the exterior solution
decays like $r^{-\ell-1}$. For large $l$, it may not be possible to capture this behaviour accurately within the infinite element basis which typically includes only inverse powers of $r$ up to a relatively low polynomial degree. This issue can
be mitigated by including part of the exterior domain within the standard FEM discretisation \citep[e.g.][]{gharti2018spectral}, but at the cost of adding proportionally many degrees of freedom into the problem. 

Given this background, the aim of this paper is to describe methods for the implementation of the DtN and multipole methods for solving Poisson's equation within the context of general-purpose and open-source finite-element packages such as MFEM \citep[][]{anderson2021mfem}, FEniCS \citep[][]{logg2012automated}, and Firedrake \citep[][]{rathgeber2016firedrake}, and to assess the  practical viability of these methods for large-scale geophysical applications. The implementation of the large-domain method within any of these packages is straightforward, but it provides a useful point of comparison for the more complicated schemes. A discussion of the infinite element method lies beyond our current scope, with its implementation depending on non-standard finite-element discretisation that would require a substantial amount of bespoke code. Within this work we focus on the MFEM package, this being a highly extensible C++ library that provides the necessary element-level control for  efficient parallel implementations of the DtN and multipole methods. The implementation of the DtN and multipole methods using popular higher-level packages such as FEniCS and Firedrake is also discussed, and while these packages do not readily allow for the same low-level optimisations, the methods described should still be of practical use. 

It is worth commenting that finite-element methods are not the only ones available for solving Poisson's equation in laterally heterogeneous planets. For example, recent work by \cite{maitra2019non} and \cite{myhill2025} shows that pseudo-spectral methods can be combined with a referential formulation of Poisson's equation to provide highly accurate and efficient solutions in planets that do not deviate too strongly from spherical symmetry. Such methods are not, however, likely to be of use within GIA modelling. This is due to the need within the coupled viscoelastic problem to account for very strong lateral viscosity variations, with such lateral variations presenting a substantial challenge
within pseudo-spectral calculations. 

The paper is organised as follows. In \cref{sec:problem}, the numerical issue of an infinite domain is discussed under the weak formulation. From \cref{sec:DN} to \cref{sec:mp}, the implementation of the domain truncation, DtN and multipole methods are described in detail, and the methods benchmarked using a simple but non-trivial problem that admits a closed-form solution. The applications of DtN and multipole to the linearised problem relevant to coupled elastic or viscoelastic problems are described in \cref{sec:lin}. Having established the accuracy and practical viability of our DtN and multipole implementations, in \cref{sec:applications} we then further illustrate the potential of the DtN method within more realistic applications by considering the calculation of the static gravitational fields for PREM \citep[][]{dziewonski1981preliminary} and a simple model for the moon Phobos. Finally, in \cref{sec:conclusions}
we summarise the contributions of this work and comment on their potential in the context of GIA modelling.
Within the main text, we focus on Poisson's equation in $\R^3$, this being the case of direct geophysical 
interest. The methods discussed can, with suitable modification, be applied to Poisson's equation in $\R^{2}$ which 
can sometimes be useful and we provide the necessary details in Appendix \ref{app:2d}.

\section{Methods}\label{sec:methods}
\subsection{Poisson equation on the whole space}\label{sec:problem}

\subsubsection{Statement of the problem}
We consider the gravitational potential, $\phi$, generated by a density distribution, $\rho$, supported in a bounded planet $M\subseteq\R^3$:
\begin{equation}
\label{eq:poisson}
\Delta\phi=4\pi G\rho\quad\text{in }\R^3,\qquad  \|\nabla \phi\| \to 0\ \text{as }\|\mb{x}\|\to\infty,
\end{equation}
where $\Delta = \partial_{i}\partial_{i}$ is the Laplace operator,  $\mathbf{x}$ denotes the spatial coordinate, and $G$ the gravitational constant. The density may be discontinuous on the surface, $\partial M$, of the planet, and here both the potential and its normal derivative must be continuous; the same conditions apply to any internal surfaces across which density is discontinuous. The solution to this problem is defined uniquely up to a constant whose value can be fixed by requiring that the potential also tends to zero at infinity.

\subsubsection{Integral solution}

As is well known \citep[e.g][]{kellogg1953foundations,dahlen1999theoretical}, the solution to the above Poisson equation  can be written 
\begin{equation}
\phi(\mathbf{x}) =  4\pi G\int_{M} \Phi(\mathbf{x},\mathbf{x}') \rho(\mathbf{x}') \dif \mathbf{x}',
\label{eq:Newton}
\end{equation}
where the fundamental solution for Poisson's equation on $\R^3$ is defined by 
\begin{equation}
    \Phi(\mathbf{x},\mathbf{x}') =  \frac{-1}{4\pi \|\mathbf{x}-\mathbf{x}'\|}.
\end{equation}
This integral solution can be used as a basis for numerical calculations \citep[e.g.][]{latychev2005glacial}, but doing so requires a careful handling of the singular integrand, while the efficient parallelisation is prohibited by the non-local nature of the calculations at each point. Through an application of the 
so-called fast multipole method \citep[e.g][]{ethridge2001new, may2011optimal}, the numerical efficiency of the integral 
method can be substantially increased, but there remain issues with efficient large-scale parallelisation and in coupling the gravitational calculation 
to  finite-element discretisations for elasticity or viscoelasticity. 

\subsubsection{Weak formulation on a bounded domain}

Consider a bounded domain, $B$, such that $M \subseteq B \subset \R^3$. If $\phi$ is a solution of Poisson's equation 
in eq.(\ref{eq:poisson}), then for any suitably smooth test function, $\psi$, defined on $B$, we find via an integration by parts that
\begin{equation}
\frac{1}{4\pi G}\int_{B}\nabla\psi\cdot\nabla\phi\dif\mb{x} -\frac{1}{4\pi G}\int_{\partial B}\psi\frac{\partial \phi}{\partial n}\dif S = -\int_M \psi \rho\dif\mb{x}.
\label{eq:weak}
\end{equation}
Within the derivation of this result, we have made use of the continuity of $\phi$ and its normal derivatives across $\partial M$ or any  internal boundaries. If, conversely, we ask that eq.(\ref{eq:weak}) holds for all test functions, and we further assume that $\phi$ is sufficiently regular, then it follows that $\phi$ satisfies the desired Poisson's equation within $B$. Importantly, however, the above requirement imposes no conditions on the behaviour of $\phi$ on the boundary $\partial B$, nor 
within the exterior domain, $\R^{3}\setminus B$. Nevertheless, we can take eq.(\ref{eq:weak}) as a starting point
for the various weak formulations of Poisson's equation considered below. Indeed, these methods differ precisely by the manner in which the potential on the boundary, $\partial B$, is specified or determined. 

\begin{figure}
\centering
\begin{subfigure}[b]{0.48\textwidth}
  \centering
  \begin{tikzpicture}
    \node[anchor=south west, inner sep=0] (img) at (0,0)
      {\includegraphics[width=\linewidth,clip,trim=6mm 8mm 6mm 10mm]{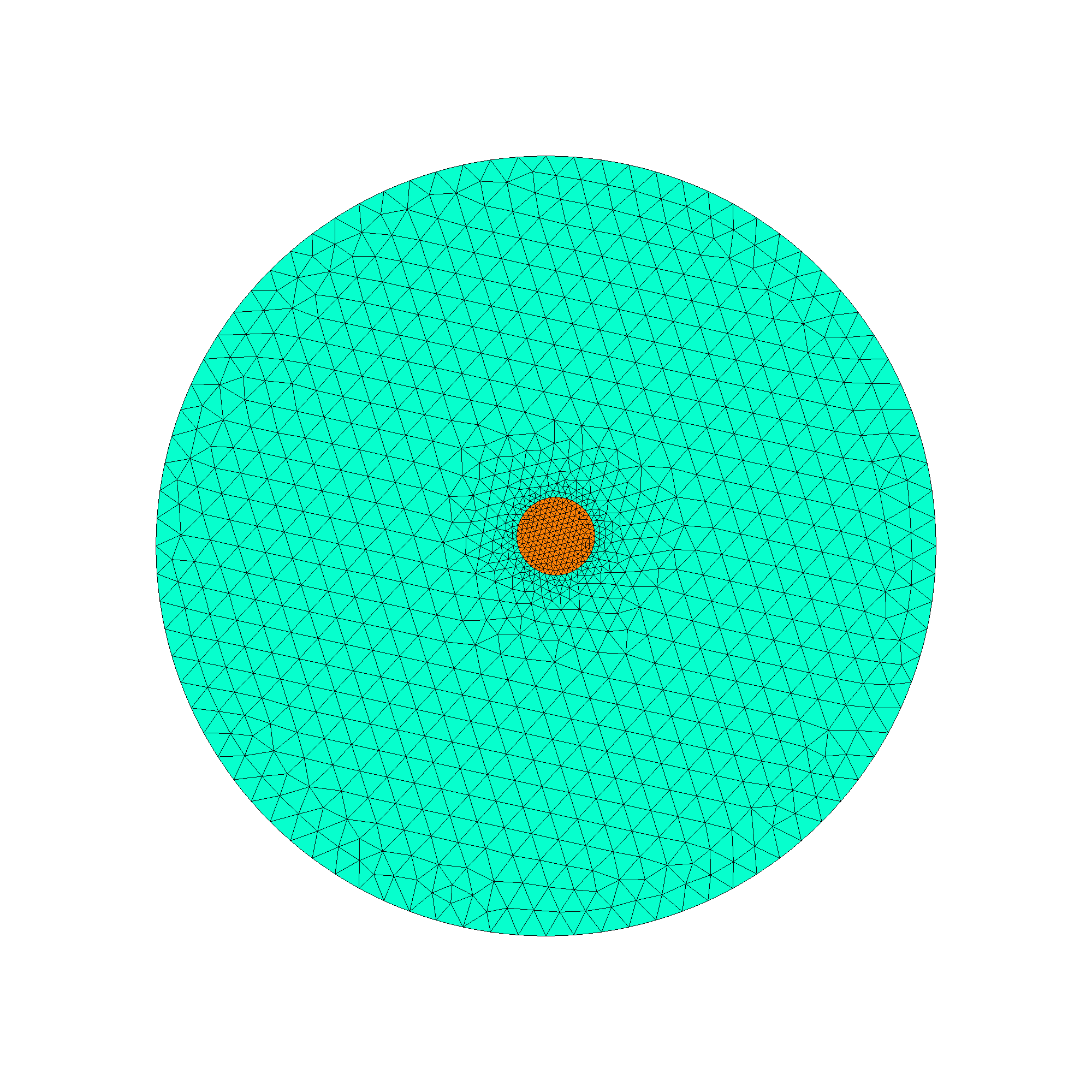}};
    \begin{scope}[x={(img.south east)}, y={(img.north west)}]
      \node[anchor=north west,font=\large\bfseries] at (0,1) {(a)};
      \draw[ann] (0.55,0.9) -- (0.45,0.78);
      \node[anchor=south west, font=\small] at (0.52,0.9) {large domain};
      \node[anchor=south west, font=\small] at (0.7,0.8) {$\phi=0$ or $\partial_n\phi=0$};
    \end{scope}
  \end{tikzpicture}
\end{subfigure}\hfill
\begin{subfigure}[b]{0.48\textwidth}
  \centering
  \begin{tikzpicture}
    \node[anchor=south west, inner sep=0] (img) at (0,0)
      {\includegraphics[width=\linewidth,clip,trim=6mm 8mm 6mm 10mm]{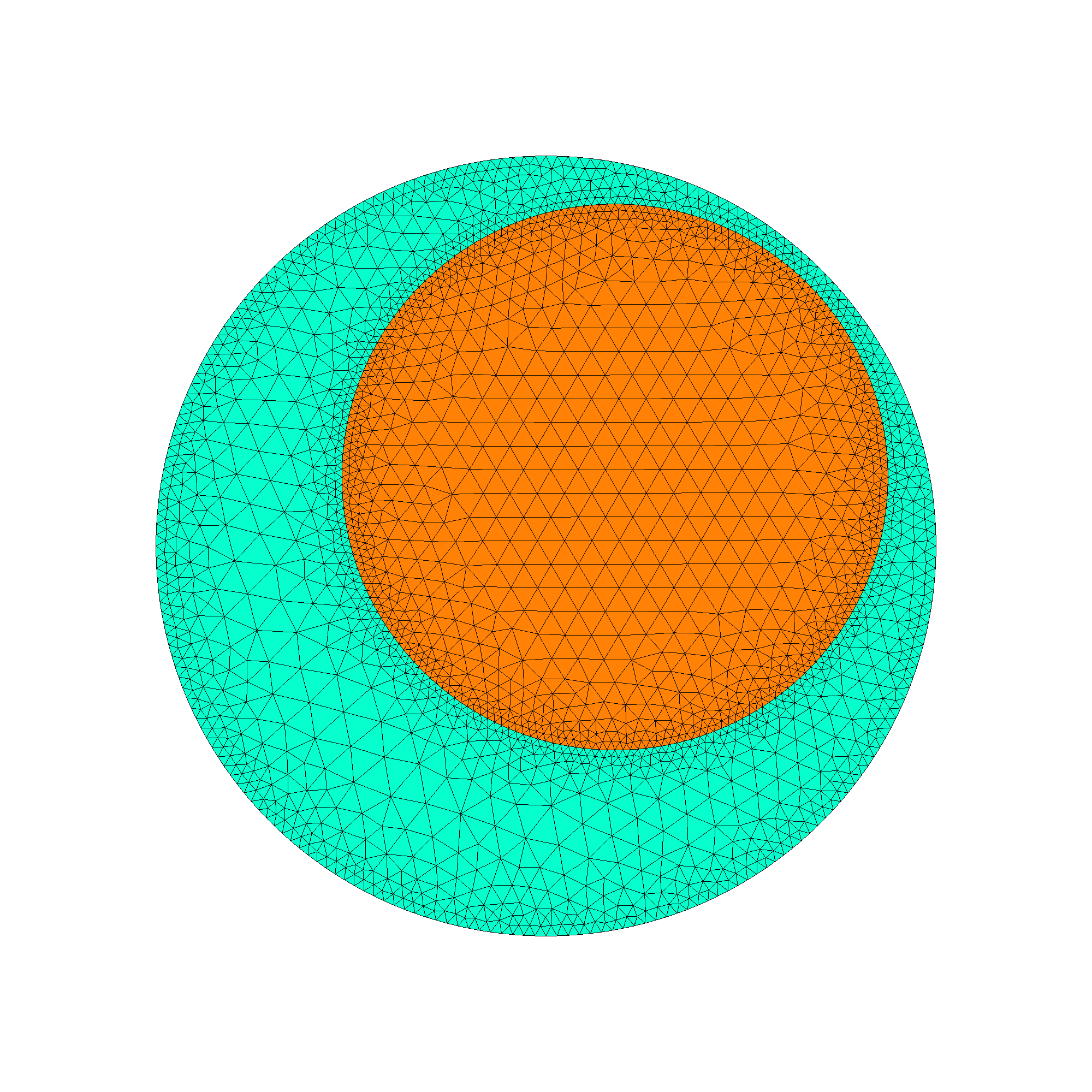}};
    \begin{scope}[x={(img.south east)}, y={(img.north west)}]

      \node[anchor=north west,font=\large\bfseries] at (0,1) {(b)};
      \draw[ann] (0.72,0.84) -- (0.61,0.72);
      \node[anchor=south west, font=\small] at (0.68,0.84) {mass distribution};
      \draw[ann] (0.89,0.29) -- (0.79,0.37);
      \node[anchor=south east, font=\small] at (1.05,0.23) {buffer layer};
      \node[anchor=south east, font=\small] at (1.09,0.64) {link $\partial_n\phi$ to $\phi_{ext}$};
    \end{scope}
  \end{tikzpicture}
\end{subfigure}
\caption{Representative meshes for the offset-sphere configuration. (a) Large-domain truncation.
(b) Buffer-layer design for the DtN and multipole methods.}
\label{fig:mesh}
\end{figure}
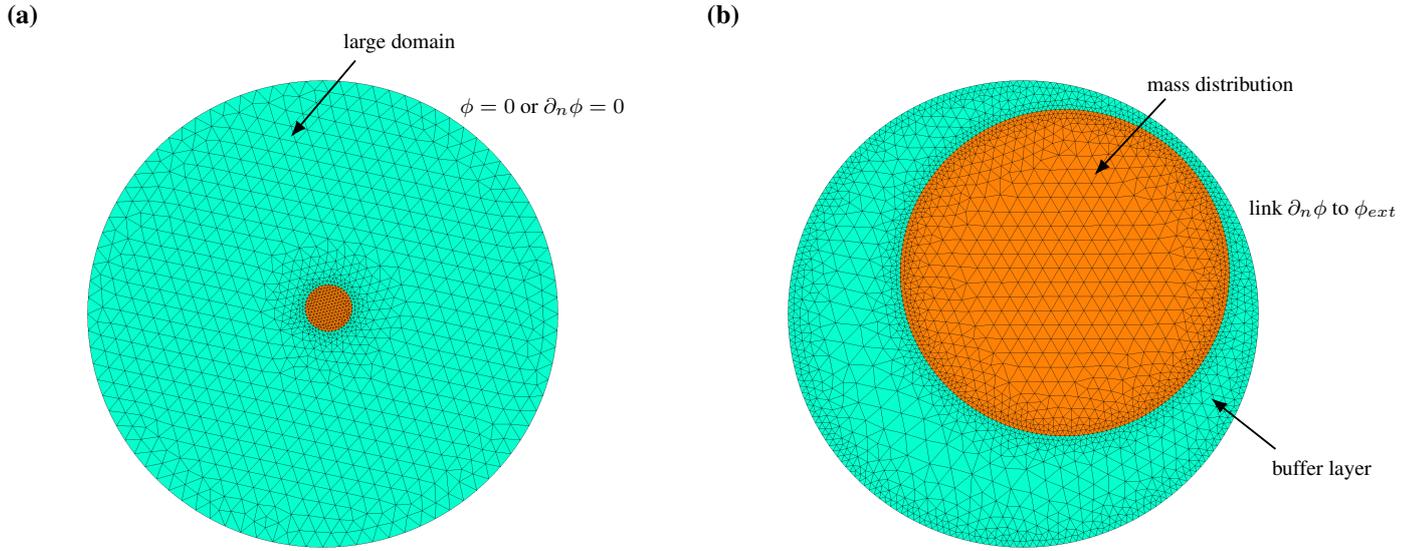

\subsection{Domain truncation}\label{sec:DN}
\subsubsection{Theory}

The integral solution in eq.(\ref{eq:Newton}) implies that both the potential and its normal derivative 
decrease to zero with increasing distance from $M$. This suggests that a simple manner for dealing with the exterior domain is to choose our finite computational domain, $B$, to be sufficiently large, and then apply 
homogeneous Dirichlet or Neumann conditions on $\partial B$. In either case the weak formulation of the problem is reduced to 
\begin{equation}
\frac{1}{4\pi G}\int_{B}\nabla\psi\cdot\nabla\phi\dif\mb{x} = -\int_M \psi\rho\dif\mb{x},
\label{eq:weak_DN}
\end{equation}
though note that for the Dirichlet case the same homogeneous boundary conditions must also be imposed on the test functions. 

It can be shown that the Dirichlet problem admits a unique solution. In the Neumann case, however, it is clear that the solution is only defined up to the addition of a constant potential. Moreover, 
if in the Neumann case we take $\psi$ equal to a non-zero constant, then
eq.(\ref{eq:weak_DN}) implies that the density must satisfy the compatibility condition
\begin{equation}
    \int_M \rho\dif\mb{x} = 0, 
    \label{eq:compat}
\end{equation}
in order for a solution of the problem to exist. This condition on the density is, of course, an entirely artificial consequence of truncating the problem to a bounded domain. For some applications this compatibility condition (or a related one within linearised calculations) is automatically met. For example, when modelling gravity anomalies it is natural to suppose that the specified density perturbations leave the total mass fixed. When the compatibility condition is not met, the use of homogeneous Dirichlet conditions is the only option within applications  of the domain truncation method. 

\subsubsection{Numerical implementation}
When using homogeneous Dirichlet conditions, the weak formulation in eq.(\ref{eq:weak_DN}) can be discretised using a standard continuous Galerkin scheme defined relative to  a finite-element mesh. The details of such discretisations are handled automatically within modern open-source FEM libraries such as MFEM, FEniCS and Firedrake. The result is a sparse, 
symmetric, and invertible, system of linear equations that can be solved efficiently using preconditioned
iterative methods. 

When applying homogeneous Neumann conditions, the discretisation proceeds in a near identical manner to arrive at  a sparse and symmetric linear system of equations. Here, however, the linear system is rank-deficient for the reasons discussed previously. Nevertheless, so long as
the density is compatible with the discretised form of eq.(\ref{eq:compat}), the linear system admits a solution (defined up to the addition of a constant potential) that can again be obtained efficiently using preconditioned iterative methods. In this case, it is necessary to project orthogonally to the null space at each stage of the Krylov iterations, but the cost 
of this additional step is trivial. Within MFEM, this can be done using their \texttt{mfem::OrthoSolver} class, while more generally methods like PETSc's \texttt{MatSetNullSpace} can be applied. 

Within modern FEM libraries, the assembly of the linear system can be performed 
very efficiently in parallel. This is because the finite-element mesh is split between available processors and  
the assembly proceeds independently on each one. Moreover, the necessary sharing of information between processors within the iterative solution of the linear systems can
be handled in a highly efficient manner by libraries such as PETSc \citep[][]{petsc-web-page} or Hypre \citep[][]{falgout2002hypre}. A key point is that processors only need to exchange data linked to their shared boundaries. Moreover, this transfer of information can be performed largely asynchronously during which time calculations linked to interior degrees of freedom are performed. 
The result is that the discretisation and solution of Poisson's equation on a bounded domain with simple boundary conditions can be achieved using modern FEM libraries in a near optimal manner. 

For later reference, it will be useful to consider a simple quantitative model for the MPI communication costs associated with each sparse matrix multiplication within 
the FEM solution of Poisson's equation. Let $N$ denote the total degrees of freedom within the finite-element discretisation, and $P$ the number of processors available. Assuming  an approximately equal partition of the domain between the processors, the number of degrees of freedom per processor is then $N/P$. Within a sparse matrix multiplication, it is necessary for processors to share only boundary data with their neighbours, and hence for 3D problems the volume of data transferred per processor can be modelled as
\begin{equation}
    D_{\mathrm{diff}} \approx k \left(\frac{N}{P}\right)^{2/3}, 
\end{equation}
where $k$ is a small constant reflecting the number of neighbours per processor (typically between 6 and 12 for 3D problems). The time required to transfer the necessary information
can then be written as
\begin{equation}
    T_{\mathrm{diff}} \approx \alpha k  + \beta k \left(\frac{N}{P}\right)^{2/3}, 
\end{equation}
where $\alpha$ represents the network latency (i.e., the time to  initiate communication between two processors), and $\beta$ is the time to transfer a unit of data. So long as the mesh partition is not made too fine, $N/P$ will be large and the second term dominates. We then have perfect weak scaling, with the communication time depending only on the ratio of $N/P$. Similarly, the number of floating point operations done per processor scales with $N/P$, and so the whole process works very well for large problems with a correspondingly large number of processors. If the problem size is fixed, and so $N$ constant, then so long as $P$ does not grow too large, we see that the communication 
time decreases as $P^{-2/3}$ and hence there is also favourable strong scaling behaviour.

\subsubsection{Benchmarks}
\label{sec:DNBench}
To illustrate the domain truncation method, we consider the application of its Dirichlet variant to the configuration shown in Fig.\ref{fig:mesh}. Here $M$ is a sphere of radius, $a$, that has constant density, $\rho_{0}$, while $B$ is the larger sphere of radius $b$. The centre of $M$ is offset from that of $B$ by a fixed distance; this choice is motivated through consideration of the DtN and multipole methods discussed below. Note that because the compatibility condition in eq.(\ref{eq:compat}) is not satisfied, a homogeneous Neumann condition cannot be applied for this problem. Given this geometry, the exact solution of Poisson's equation takes the simple form
\begin{equation}\label{eq:ana}
\phi_{\text{ex}}(r) =
\begin{cases}
-2\pi G \rho_0 \left( a^{2} - \dfrac{r^{2}}{3} \right), & r \le a, \\[5pt]
-\dfrac{4\pi G \rho_0 a^{3}}{3r}, & r > a.
\end{cases}
\end{equation}
where $r$ is the radius defined relative to the centre of $M$. By numerically solving this problem for different values of the ratio $b/a$,  we can assess the ability of the domain truncation method to approximate the exact solution.

Within our numerical calculations, tetrahedral finite-element meshes were generated using gmsh \citep[][]{geuzaine2009gmsh}. Representative examples are shown within Fig.\ref{fig:mesh}, though we note that the meshes shown are two-dimensional to aid visualisation. The boundary, $\partial M$, across which the density is discontinuous, is modelled explicitly. Second-order elements were used for the meshes to better account for curved boundaries, while element sizes
are refined close to $\partial M$; both these steps are needed to reduce discretisation errors. 
The weak form of Poisson's equation was discretised within MFEM using a continuous Galerkin scheme with third-order Lagrange elements. The resulting linear system was solved with the preconditioned conjugate gradient method using the \texttt{boomerAMG} algebraic multigrid preconditioner \citep[][]{yang2002boomeramg} provided by the Hypre library. All calculations were performed in parallel using 8 CPUs.

\begin{figure}
\centering
\begin{subfigure}{0.85\textwidth}
  \centering
  \begin{minipage}{0.49\textwidth}
    \centering
    \begin{tikzpicture}
      \node[anchor=south west, inner sep=0] (imgb) at (0,0)
        {\includegraphics[width=\linewidth]{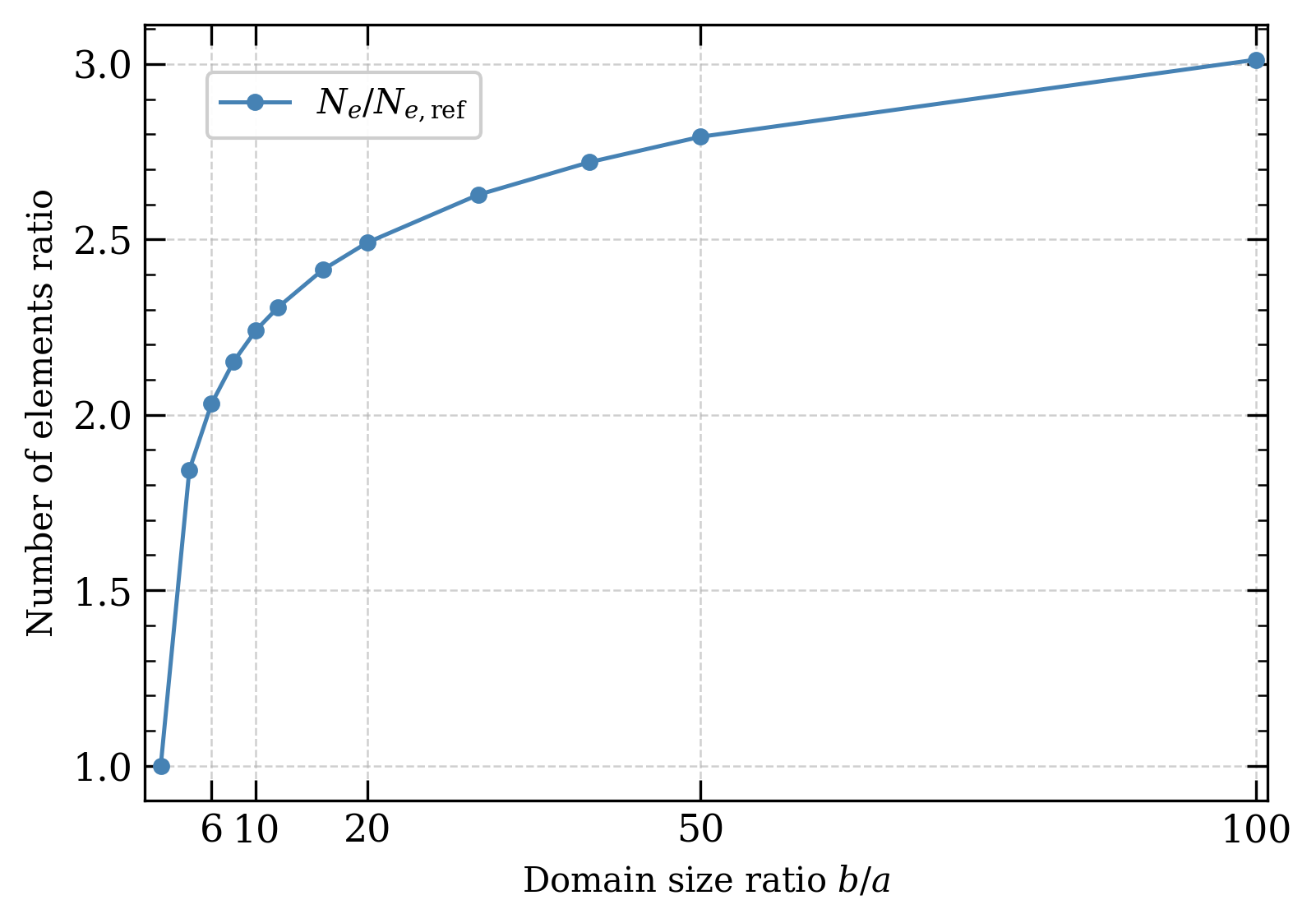}};
      \begin{scope}[x={(imgb.south east)}, y={(imgb.north west)}]
        \node[anchor=north west, font=\large\bfseries] at (-0.05,1) {(a)};
      \end{scope}
    \end{tikzpicture}
  \end{minipage}\hfill
  \begin{minipage}{0.49\textwidth}
    \centering
    \begin{tikzpicture}
      \node[anchor=south west, inner sep=0] (imgc) at (0,0)
        {\includegraphics[width=\linewidth]{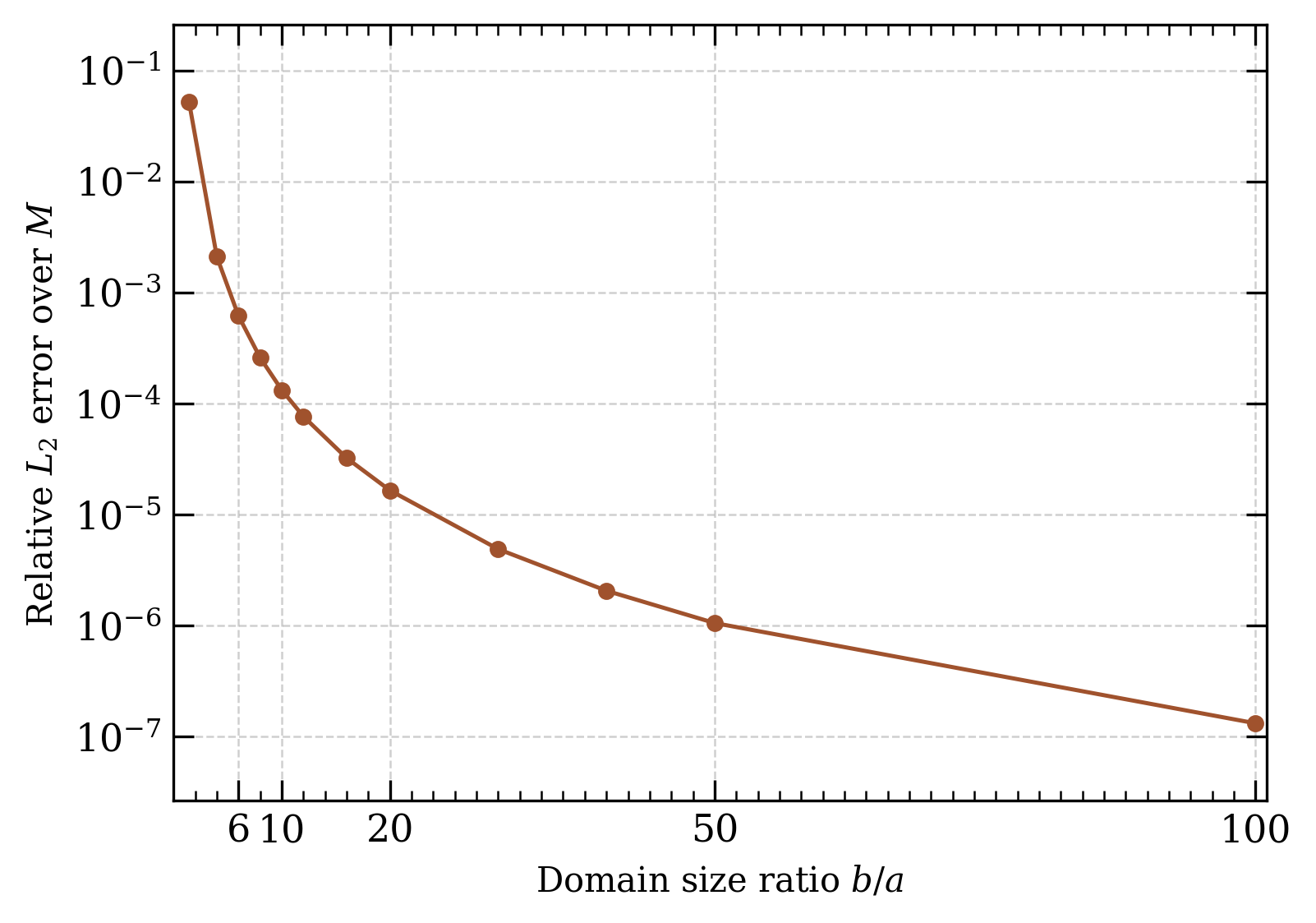}};
      \begin{scope}[x={(imgc.south east)}, y={(imgc.north west)}]
        \node[anchor=north west, font=\large\bfseries] at (-0.025,1) {(b)};
      \end{scope}
    \end{tikzpicture}
  \end{minipage}
\end{subfigure}

\caption{Large-domain truncation with the zero Dirichlet boundary condition within the offset-sphere configuration. (a) Number of tetrahedral elements normalised by that at the reference case $b/a=10/7$, $N_{e,\mathrm{ref}}=498525$. (b) Relative \(L_2\) error over $M$, $\varepsilon_{L^2(M)}$, versus \(b/a\).}
\label{fig:nd_methods}
\end{figure}

Fig.\ref{fig:nd_methods} summarises the application of the domain truncation method with Dirichlet  boundary conditions
to our benchmark problem for different values of $b/a$. In Fig.\ref{fig:nd_methods}(a) we show the number of elements as a function of $b/a$, while in (b)  the relative $L^{2}$-error within the interior domain, $M$, is shown, this being defined  by
\begin{equation}
\varepsilon_{L^{2}(M)} = \frac{\|\phi-\phi_{\text{ex}}\|_{L^{2}(M)}}{\|\phi_{\text{ex}}\|_{L^{2}(M)}}.
\end{equation}
From Fig.~\ref{fig:nd_methods}(a), as the domain size increases, the number of elements initially grows rapidly, but the rate of increase soon becomes restrained; when \( b/a \) increases from $10/7$ to $100$, the total number of elements merely triples. Because only a relatively mild mesh-size coarsening strategy is employed, this result indicates that the large-domain truncation approach is generally not limited by an explosion in the number of degrees of freedom. Meanwhile, Fig.~\ref{fig:nd_methods}(b) shows that the \( L^2(M) \) error decreases steadily with increasing \( b/a \), and a domain size of \( b/a = 50 \) is required to achieve approximately \( \varepsilon_{L^{2}(M)} \approx 10^{-6} \). Although the computational cost remains acceptable according to Fig.~\ref{fig:nd_methods}(a), adopting a computational domain fifty times larger than the planet introduces additional complexity in numerical implementation. The exact solver time required to reach \( \varepsilon_{L^2(M)} = 1.06 \times 10^{-6} \) at \( b/a = 50 \), namely \( T_{\mathrm{ref}} = 56.01\,\mathrm{s} \), is therefore taken as the reference for evaluating the performance of the DtN and multipole methods presented below.


\subsection{Dirichlet-to-Neumann (DtN) method}\label{sec:dtn}
\subsubsection{Theory}

We now assume that the computational domain, $B\subseteq \R^3$, is a ball of radius, $b$. Within the exterior domain, $\R^3\setminus B$, the potential is harmonic, and so admits the well-known representation
\begin{equation}
    \label{eq:ext}
    \phi(r,\theta,\varphi) = \sum_{\ell m} \left(\frac{b}{r}\right)^{\ell+1} \phi_{\ell m}(b) Y_{\ell m}(\theta,\varphi).
\end{equation}
Here  $(r,\theta,\varphi)$ denote spherical polar co-ordinates defined relative to the centre of $B$, 
$Y_{\ell m}$ is the fully normalised real spherical harmonics \citep[e.g.][]{dahlen1999theoretical} of degree, $\ell$, and order, $m$, and 
$\phi_{\ell m}(b)$ is the $(\ell,m)_{\mathrm{th}}$ spherical coefficient of $\phi$ on $\partial B$ that can be written as
\begin{equation}
    \phi_{\ell m}(b) = \frac{1}{b^{2}}\int_{\partial B} \phi\,Y_{\ell m} \dif S.
\end{equation}
Taking the normal derivative of the exterior solution (\ref{eq:ext}) at the spherical boundary $r=b$ yields
\begin{equation}
\left.\frac{\partial \phi}{\partial n}\right|_{\partial B} = -\sum_{\ell m}\frac{\ell+1}{b}\,\phi_{\ell m}(b)\,Y_{\ell m}(\theta,\varphi).
\end{equation}
This relation shows how the values of the potential on $\partial B$ (the Dirichlet data) can be 
mapped linearly into those for the normal derivative (the Neumann data), and hence this is known 
as a Dirichlet to Neumann (DtN) operator. DtN operators can also be defined on more general surfaces that 
separate interior and exterior domains, but here we see the very simple diagonal form this operator takes 
when defined on a  spherical boundary and expressed relative to the spherical harmonic basis. 

Returning to eq.(\ref{eq:weak}), we can use the DtN operator to express the required normal derivative on $\partial B$, yielding
\begin{equation}
\label{eq:weak_DtN}
   \frac{1}{4\pi G} \int_B\nabla \psi\cdot\nabla \phi \dif \mathbf{x} + \frac{1}{4\pi G}\sum_{\ell m}(\ell+1)b\,\psi_{\ell m}(b)\phi_{\ell m}(b) = -\int_{M} \psi \rho \dif\mathbf{x}.
\end{equation}
If this equation holds for all test functions, $\psi$,  we see that $\phi$ satisfies the required Poisson equation within $B$, while its normal derivative on $\partial B$ is given by the results of the DtN operator acting on $\phi$ at the boundary. Eq.(\ref{eq:weak_DtN}) therefore provides a suitable weak formulation of the Poisson equation that accounts exactly for the exterior domain without the need for it to be modelled explicitly.

\subsubsection{Numerical implementation}
The weak formulation for Poisson's equation given in eq.(\ref{eq:weak_DtN}) differs from that discussed previously within the context of the domain truncation method only through the  addition of the DtN term which is represented through a symmetric bilinear form
\begin{equation}
\label{eq:DtNform}
    (\psi, \phi) \mapsto \frac{1}{4\pi G}\sum_{\ell m}(\ell+1)b\,\psi_{\ell m}(b)\phi_{\ell m}(b), 
\end{equation}
and hence we can focus on the implementation of this new term alone.  

Within numerical work it is, of course, necessary to truncate the spherical harmonic expansion to some degree, $\ell_{\max}$. From eq.(\ref{eq:ext}), we know that 
the degree-$\ell$ component of the exterior solution decays as $r^{-\ell-1}$. Sufficient accuracy within a given problem can, therefore, be achieved by (i) increasing $\ell_{\max}$
while holding $b$ fixed, (ii) increasing $b$ for fixed $\ell_{\max}$, or (iii) some combination of the two preceding techniques. As will be seen below, increasing $\ell_{\max}$
requires additional non-local communication that can affect efficient parallelisation, whereas increasing $b$ increases the degrees of freedom. 

The DtN form in eq.(\ref{eq:DtNform}) is comprised of spherical harmonic coefficients of the trial and test functions, with each such coefficient being expressed as an integral over $\partial B$. This has the effect of coupling all degrees of freedom within the finite-element discretisation linked to the exterior boundary, $\partial B$. It is this property that presents the main challenge to the efficient parallel implementation of the DtN form. In particular, were this term to be fully assembled as a matrix, it would contain a large and dense sub-block associated with $\partial B$. Moreover, the definition of the DtN form as a  sum over spherical harmonic coefficients is not natively supported within FEM libraries such as MFEM, FEniCS, and Firedrake, and hence its implementation requires additional work. To proceed within the MFEM framework, the discretised DtN form is defined as a custom class that inherits from \texttt{mfem::Operator}, the latter being the abstract base class for linear operators within the library. To understand this construction, it is helpful to begin with the serial implementation. Relative to a finite-element discretisation, the DtN form corresponds to a matrix, $\mathbf{B}$, that 
can be factored as
\begin{equation}
    \mathbf{B} = \mathbf{C}^{T}\mathbf{C}. 
\end{equation}
Here, $\mathbf{C}$ maps a discretised scalar field, $\phi$, to a vector of its spherical harmonic coefficients up to degree, $\ell_{\max}$, and 
then applies the degree-dependent scaling $\phi_{\ell m} \mapsto \sqrt{\frac{(\ell+1) b}{4\pi G}}\phi_{\ell m}$. The matrix, $\mathbf{C}$, is pre-assembled as a sparse 
matrix with non-zero contributions only from boundary elements. The necessary integrals are approximated using a standard quadrature scheme, while trigonometric  functions and associated Legendre polynomials are efficiently computed  using stable recursion schemes. The action of $\mathbf{B}$ on a vector, $\mathbf{x}$, requires just two sparse products, with the small intermediate vector, $\mathbf{y} = \mathbf{C}\mathbf{x}$,
being  stored as a mutable  class member. 

To parallelise this process, we form a split MPI communicator that comprises only processors that contain part of the boundary, $\partial B$. On each boundary processor we assemble
a sparse matrix, $\mathbf{C}$, that is defined as before, but now only accounts for contributions from that processor's portion of the boundary. The action of the overall operator, $\mathbf{B}$, corresponding to the DtN form then proceeds in three stages. First, each boundary processor acts its sparse matrix, 
$\mathbf{C}$, on its part of the vector. Next, a single call to \texttt{MPI\_Allreduce} is  made to sum the contributions from each boundary processor to form  the suitably weighted vector of spherical harmonic coefficients. Finally, each processor acts $\mathbf{C}^{T}$ on the coefficient vector to obtain its part of the output vector. 

Within the above implementation, the assembly of the DtN term is done locally on each processor, and hence we  expect perfect weak scaling with the number of processors. During the action of the operator on a vector, the data shared between all the boundary processors depends only on $\ell_{\max}$, and hence we can set
\begin{equation}
    D_{\mathrm{DtN}} \approx (\ell_{\max}+1)^{2}. 
\end{equation}
If $P_{b}$ denotes the number of boundary processors, then the communication time can be written as
\begin{equation}
    T_{\mathrm{DtN}} \approx \alpha \log_{2}( P_{b}) + \beta (\ell_{\max}+1)^{2}, 
\end{equation}
where the dependence on $\log_{2}P_{b}$ arises from the efficient implementation of \texttt{MPI\_Allreduce} using a binary doubling algorithm. Assuming a roughly even partition of 
the mesh between processors, we would expect that $P_{b} \approx P^{2/3}$, and hence we can also write 
\begin{equation}
    T_{\mathrm{DtN}} \approx \frac{2}{3}\alpha \log_{2}(P) + \beta (\ell_{\max}+1)^{2}. 
\end{equation}
So long as $\ell_{\max}$ is not too large, $T_{\mathrm{DtN}}$ will be dominated by latency, and recalling our earlier analysis of the diffusion term within Poisson's equation we find
\begin{equation}
    \frac{T_{\mathrm{DtN}} }{T_{\mathrm{diff}}} \approx \frac{2 \alpha}{3 \beta k }\log_{2}(P) \left(\frac{P}{N}\right)^{2/3}.
\end{equation}
In terms of strong scaling,  we see that the ratio of the communication times grows with the number of processors as $\log_{2}(P) P^{2/3}$, though this is at least sub-linear for large $P$. The weak scaling behaviour is better, however, growing only as $\log_{2}(P)$. Either way, this analysis makes clear the issue with even a careful implementation of the DtN term; as the number of processors grows, there will be a point at which the total communication time is dominated by sharing 
spherical harmonic coefficients between the boundary processors. Within this analysis, it is notable that the ratio of the communication times is independent of 
$\ell_{\max}$, but the number of floating point operations per boundary processor linked to the discretised DtN form scales as
\begin{equation}
(\ell_{\max}+1)^{2} \left(\frac{N}{P}\right)^{2/3}, 
\end{equation}
and hence we  still expect that the total cost for acting the operator on a vector will rise with $\ell_{\max}$. 

Within the libraries FEniCS and Firedrake, the element-level access required to implement the DtN form as described above is not readily available. Nevertheless, a workable approach can be developed. For degree, $\ell$, and order, $m$, we can consider the linear form
\begin{equation}
\phi \mapsto \sqrt{\frac{(\ell+1) }{4\pi G b^3}}\int_{\partial B}  Y_{\ell m} \phi \dif S. 
\end{equation}
Relative to a finite-element discretisation, such a form is represented by a vector that we write as $\mathbf{c}_{\ell m}$ and that can readily be defined and assembled using the unified form language of  FEniCS and Firedrake. The matrix corresponding to the DtN form can then be written as a sum of tensor products
\begin{equation}
    \mathbf{B} = \sum_{\ell m} \mathbf{c}_{\ell m} \otimes \mathbf{c}_{\ell m},
\end{equation}
which is taken up to the truncation degree, $\ell_{\max}$. This matrix should not, of course, be constructed explicitly. Rather, 
using a method such  as \texttt{MatCreateLRC} from the PETsc library, the action of $\mathbf{B}$ can be combined with that of $\mathbf{A}$ in an efficient matrix-free manner. This implementation of the DtN term is unlikely to be as efficient as the approach with MFEM detailed above, though the overall form of the parallel scaling should be identical. 

The full linear system associated with eq.(\ref{eq:weak_DtN}) can be written schematically in the form
\begin{equation}
    (\mathbf{A} + \mathbf{B})\mathbf{x} = \mathbf{y}, 
\end{equation}
where $\mathbf{A}$ is a sparse matrix associated with the bilinear form
\begin{equation}
    (\phi,\psi) \mapsto  \frac{1}{4\pi G} \int_B\nabla \psi\cdot\nabla \phi \dif \mathbf{x}, 
\end{equation}
$\mathbf{B}$ is the matrix corresponding to the DtN term, $\mathbf{x}$ represents the discretisation of the potential, and $\mathbf{y}$ that of the force term within the equation. It is not necessary nor desirable to explicitly  add these matrices, with their sum instead being implemented lazily. Both $\mathbf{A}$ and $\mathbf{B}$ are symmetric, 
and so we can still apply the preconditioned conjugate gradient method. To form a suitable preconditioner, we cannot readily make use of the full matrix $\mathbf{A} +\mathbf{B}$ because the DtN term is implemented only through its action and most good preconditioners require element-level access. We have, however, found it sufficient to build the preconditioner using $\mathbf{A}$ alone. In detail, because $\mathbf{A}$ is singular, we assemble the sparse matrix corresponding to the
shifted form 
\begin{equation}
    (\psi, \phi) \mapsto  \frac{1}{4\pi G} \int_B \left(\nabla \psi\cdot\nabla \phi + \epsilon\, \psi \phi \right)\dif \mathbf{x}, 
\end{equation}
for small $\epsilon > 0$, and use this to define an algebraic multigrid preconditioner  using \texttt{boomerAMG} from the \texttt{Hypre} library. Using this approach to preconditioning, the number of iterations required to converge within the DtN problem has been found to be nearly identical to that for solving a simple Neumann problem on the same domain. This reflects the fact that the DtN term acts as both a small and low-rank perturbation to the original linear system.

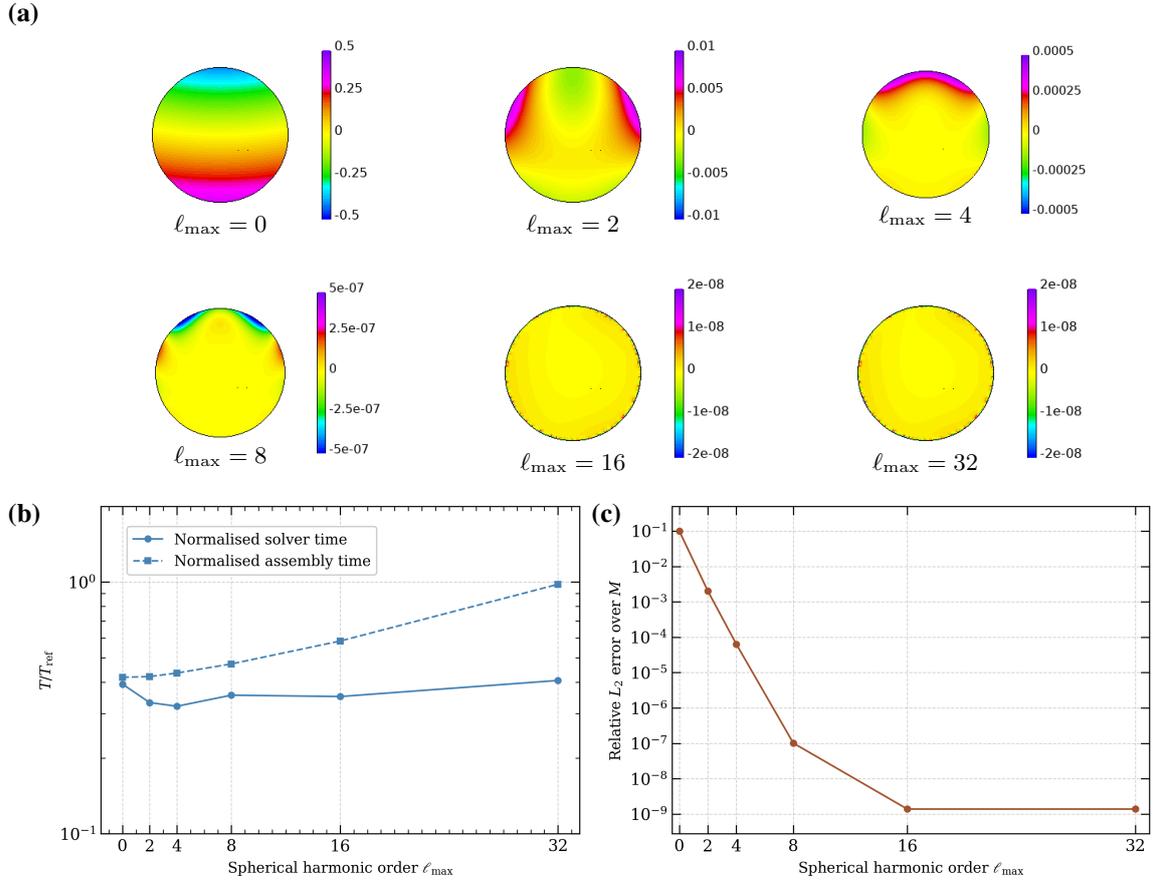
\begin{figure}
\centering
\begin{subfigure}{0.85\textwidth}
  \centering
  \setlength{\tabcolsep}{6pt}
  \renewcommand{\arraystretch}{1.0}

  \begin{tikzpicture}
    \node[anchor=south west, inner sep=0] (grid) at (0,0) {
      \begin{minipage}{\linewidth}
        \centering
        \begin{tabular}{@{}c c c@{}}
        
          \begin{minipage}{0.28\textwidth}\centering
            \begin{tikzpicture}
              \node[anchor=south west, inner sep=0] (p) at (0,0)
                {\includegraphics[width=\linewidth]{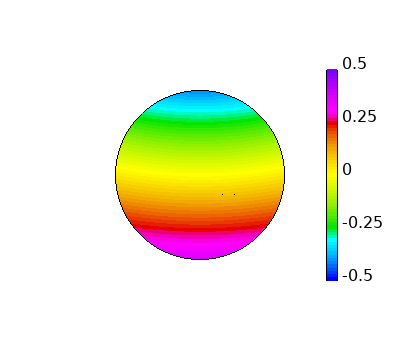}};
            \end{tikzpicture}\\[-4.5em]
            {\normalsize $\ell_{\max}=0$}
          \end{minipage} &
          \begin{minipage}{0.28\textwidth}\centering
            \begin{tikzpicture}
              \node[anchor=south west, inner sep=0] (p) at (0,0)
                {\includegraphics[width=\linewidth]{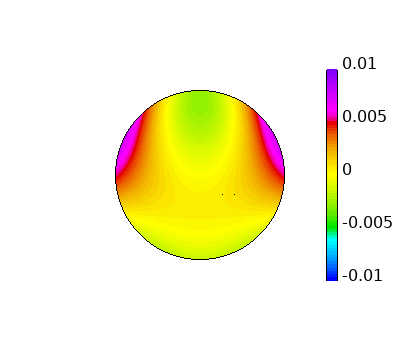}};
            \end{tikzpicture}\\[-4.5em]
            {\normalsize $\ell_{\max}=2$}
          \end{minipage} &
          \begin{minipage}{0.28\textwidth}\centering
            \begin{tikzpicture}
              \node[anchor=south west, inner sep=0] (p) at (0,0)
                {\includegraphics[width=\linewidth]{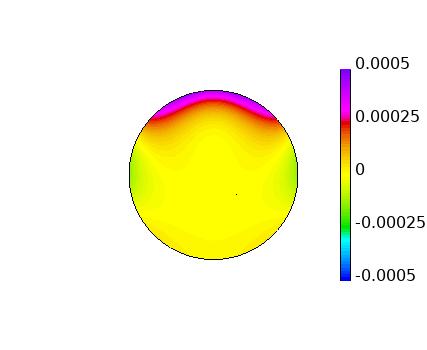}};
            \end{tikzpicture}\\[-4.5em]
            {\normalsize $\ell_{\max}=4$}
          \end{minipage}
          \\
          [0.4em]
          
          \begin{minipage}{0.28\textwidth}\centering
            \begin{tikzpicture}
              \node[anchor=south west, inner sep=0] (p) at (0,0)
                {\includegraphics[width=\linewidth]{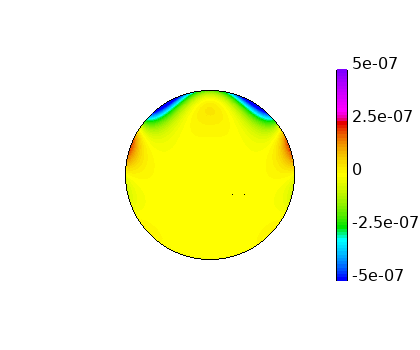}};
            \end{tikzpicture}\\[-4.5em]
            {\normalsize $\ell_{\max}=8$}
          \end{minipage} &
          \begin{minipage}{0.28\textwidth}\centering
            \begin{tikzpicture}
              \node[anchor=south west, inner sep=0] (p) at (0,0)
                {\includegraphics[width=\linewidth]{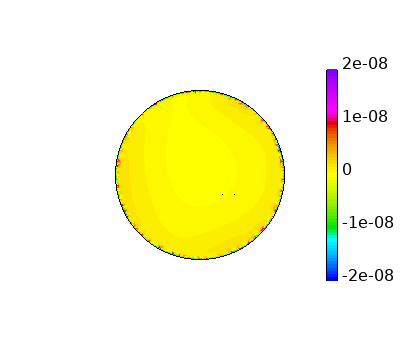}};
            \end{tikzpicture}\\[-4.5em]
            {\normalsize $\ell_{\max}=16$}
          \end{minipage} &
          \begin{minipage}{0.28\textwidth}\centering
            \begin{tikzpicture}
              \node[anchor=south west, inner sep=0] (p) at (0,0)
                {\includegraphics[width=\linewidth]{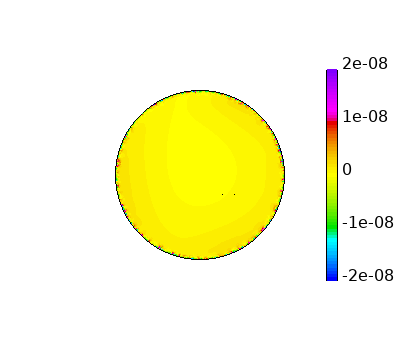}};
            \end{tikzpicture}\\[-4.5em]
            {\normalsize $\ell_{\max}=32$}
          \end{minipage}
        \end{tabular}
      \end{minipage}
    };
    
    \begin{scope}[x={(grid.south east)}, y={(grid.north west)}]
      \node[anchor=north west, font=\large\bfseries] at (0,1) {(a)};
    \end{scope}
  \end{tikzpicture}
\end{subfigure}

\vspace{1.0em}

\begin{subfigure}{0.85\textwidth}
  \centering
  \begin{minipage}{0.49\textwidth}
    \centering
    \begin{tikzpicture}
      \node[anchor=south west, inner sep=0] (imgb) at (0,0)
        {\includegraphics[width=\linewidth]{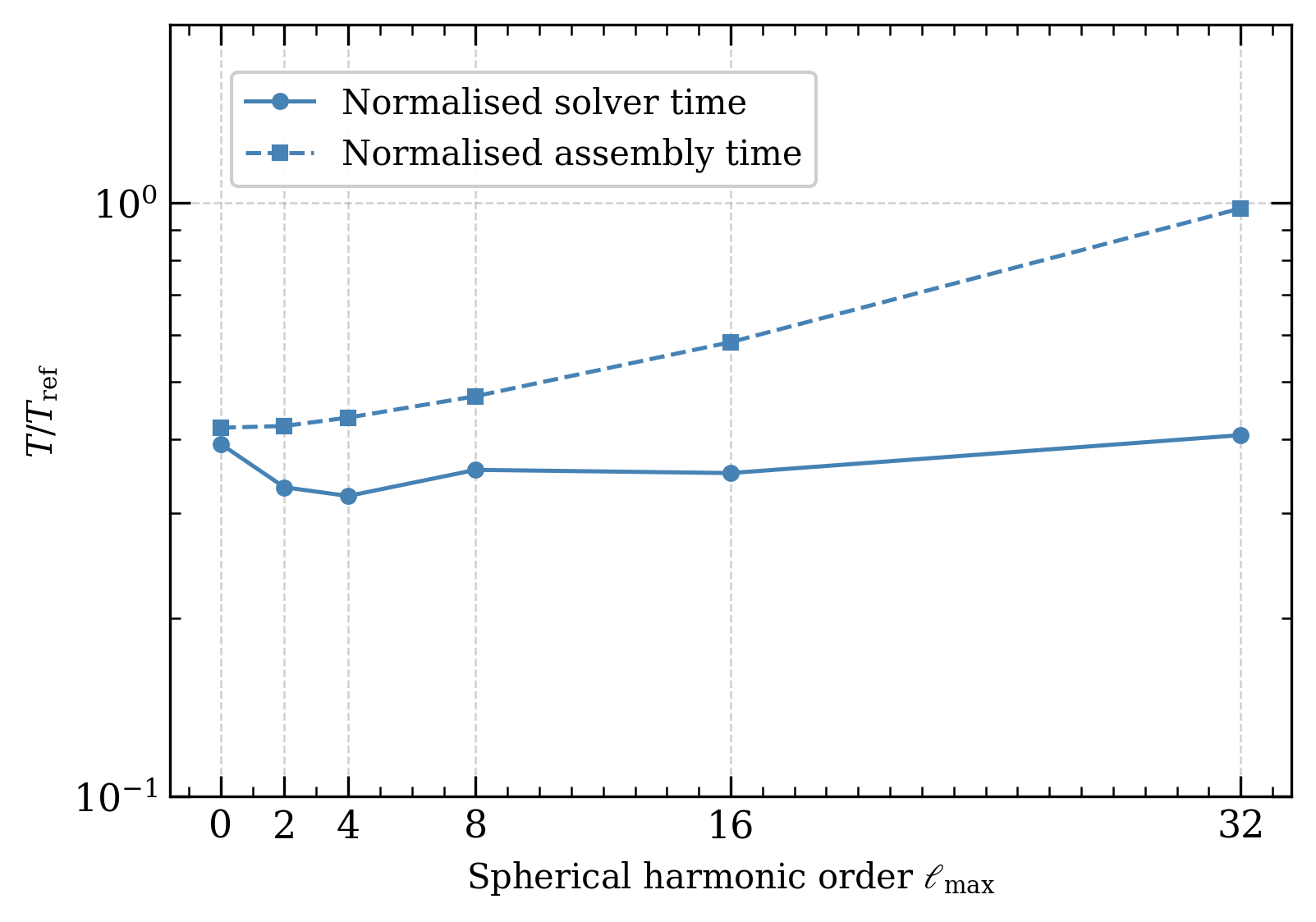}};
      \begin{scope}[x={(imgb.south east)}, y={(imgb.north west)}]
        \node[anchor=north west, font=\large\bfseries] at (-0.05,1) {(b)};
      \end{scope}
    \end{tikzpicture}
  \end{minipage}\hfill
  \begin{minipage}{0.49\textwidth}
    \centering
    \begin{tikzpicture}
      \node[anchor=south west, inner sep=0] (imgc) at (0,0)
        {\includegraphics[width=\linewidth]{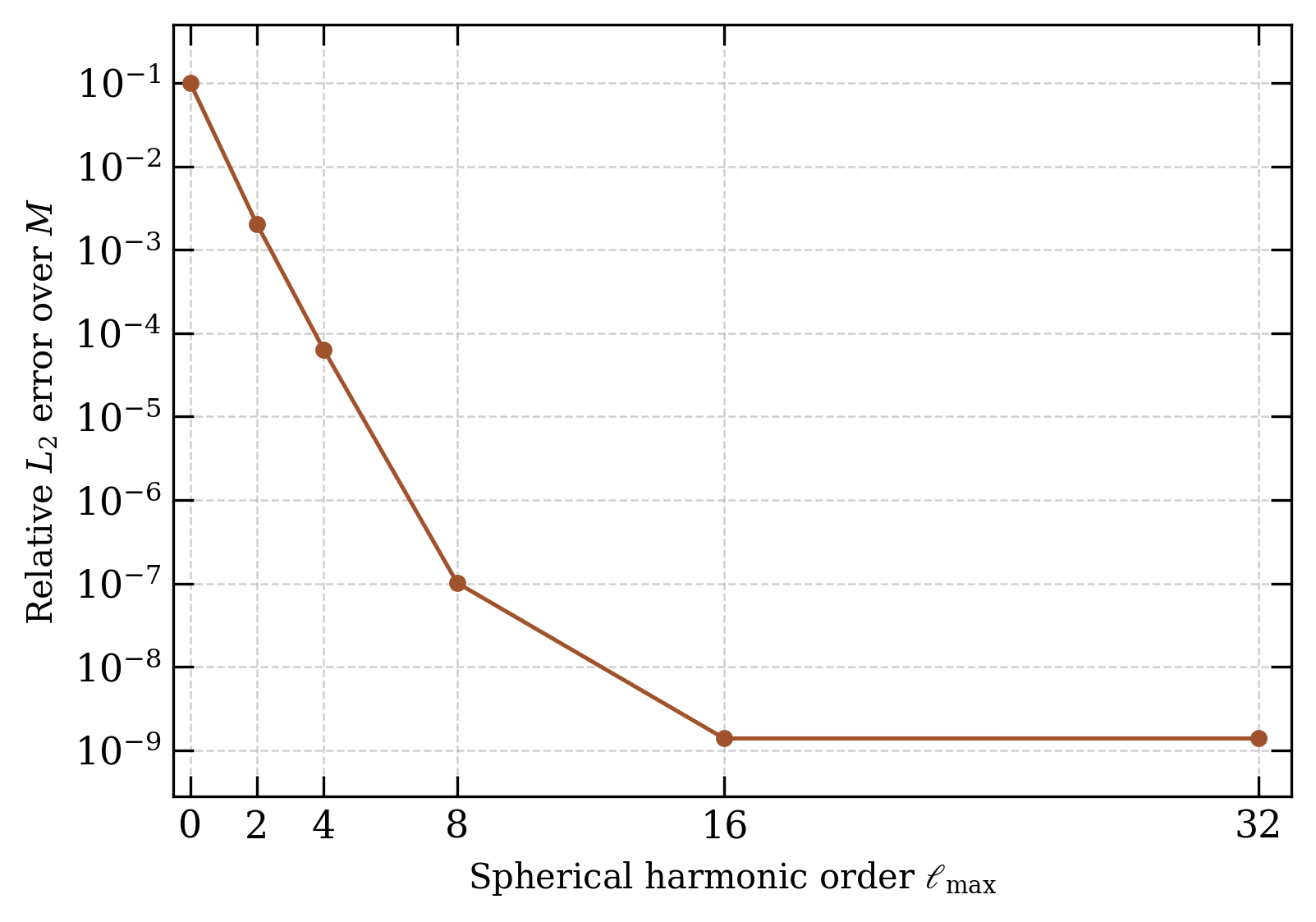}};
      \begin{scope}[x={(imgc.south east)}, y={(imgc.north west)}]
        \node[anchor=north west, font=\large\bfseries] at (-0.025,1) {(c)};
      \end{scope}
    \end{tikzpicture}
  \end{minipage}
\end{subfigure}

\caption{DtN method for the offset-sphere configuration with ${b/a=10/7}$. (a) Signed relative error distributions over $M$ for six spherical harmonic orders (\(\ell_{\max}=\{0,2,4,8,16,32\}\)). (b) Assembly time (normalised by the time for assembling the diffusion term with the zero Dirichlet condition, $T_{ref}=3.81\mathrm{s}$) and solver time (normalised by that achieving $\varepsilon_{L^2(M)}=1.06\times 10^{-6}$ with a large domain of $b/a=50$, $T_{ref}=56.01\mathrm{s}$) versus \(\ell_{\max}\). All times are averaged over 20 runs with 8 CPU processors. (c) Relative \(L_2\) error over $M$, $\varepsilon_{L^2(M)}$, versus \(\ell_{\max}\).}
\label{fig:dtn}
\end{figure}

\subsubsection{Benchmarks}

To test the performance of the DtN method, we again use the offset sphere configuration discussed within Section \ref{sec:DNBench}. Here, however, we fix the size of the computational domain to $b/a = 10/7$, 
and then vary the truncation degree used within the DtN form. As before, the finite-element discretisation was performed by MFEM using third order Lagrange elements defined on a second order tetrahedral mesh, generated by gmsh, and all calculations were run on 8 CPUs. The results are summarised within Fig.\ref{fig:dtn}, with pointwise and $L^{2}$ relative errors shown for the solution restricted to the interior domain, $M$. As can be seen, for this problem the relative error decreases very rapidly with $\ell_{max}$. By $\ell_{\max}=16$, the relative pointwise error clearly shows artifacts linked to the finite-element discretisation, and particularly around the curved boundary, $\partial M$.  This 
suggests that for such values of  $\ell_{max}$ the exterior boundary condition  on $\partial B$ is essentially exact. 

Within Fig.\ref{fig:dtn}(b), we show the timing for assembly and solution of the linear system within the DtN method, relative to that for assembling the diffusion term on the same mesh and that for achieving around $10^{-6}$ error using large-domain truncation, respectively. All times have been averaged over 20 runs. The assembly time for the DtN operator increases with $\ell_{\max}$
in an expected manner, but remains comparable to that needed for the assembly of the diffusion term. The solver time for the DtN method varies insignificantly with $\ell_{\max}$ and is always below a half of that to achieve $10^{-6}$ accuracy via large-domain truncation, whereas for the DtN method the error effortlessly reduces to $10^{-9}$ when $\ell_{\max}\geq 16$. It should be emphasised that the number of iterations required for convergence was  unchanged between solutions using Dirichlet conditions or the DtN method, and hence the solver times shown are indicative of the cost per matrix-vector product within the iterative solution. 

Clearly, the appropriate choice of $\ell_{\max}$ within the DtN method is problem dependent, and hence the values above need not to be representative. Nevertheless, this example does establish the correctness and  viability of our implementation of the DtN method within the solution of a realistic 3D problem, while the results point to some general features. First, as the truncation degree used within the DtN is increased, the accuracy of the method increases dramatically, this reflecting the fact that the spherical harmonic power of the potential decays rapidly for higher degrees within the exterior domain.  Second, the assembly time of the method increases with $\ell_{\max}$, but this is usually small relative to the solver time for realistic applications. The latter observation suggests that for the problem considered, the efficiency of the DtN method is controlled 
primarily through non-local communication between the boundary processors and not the additional floating point operations required. 
Later, in Section \ref{sec:par}, we examine the performance of the DtN method as the number of CPUs is increased. Another key feature of the DtN method is that the necessary truncation degree for a given level of accuracy can be adjusted by changing the radius, $b$,
of the computational domain. By increasing $b$, the required value of $\ell_{\max}$ can be reduced, thereby decreasing the additional overhead associated with the non-local communication reduced. This, of course, increases the total degrees of freedom within the problem, and so a balance must be found. Moreover, the mesh close to the boundary $\partial B$ needs not be as fine as within the interior domain or close to $\partial M$. In this manner, the boundary degrees of freedom and the number of boundary processors can be reduced, further improving the efficiency of the method.

\subsection{Multipole expansion method}\label{sec:mp}
\subsubsection{Theory}

As with the DtN method, we suppose that $B$ is a ball of radius $b$ that contains $M$.  The basis for this method is the following expansion 
\begin{equation}
    \Phi(\mb{x},\mb{x'})=- \frac{1}{r}\sum_{\ell m}\frac{1}{2\ell+1}\left(\frac{r'}{r}\right)^{\ell}Y_{\ell m}(\theta',\varphi')\,Y_{\ell m}(\theta,\varphi),
    \label{eq:fund_sh}
\end{equation}
for the fundamental solution of Poisson's equation in $\R^3$. Here the spherical polar co-ordinates of $\mathbf{x}$ are denoted by $(r,\theta,\varphi)$,
and those for $\mathbf{x}'$ are denoted by $(r',\theta',\varphi')$, while it is assumed that $r > r'$ \citep[e.g.][]{dahlen1999theoretical}.
Combing this expansion with the integral solution for the problem in eq.(\ref{eq:Newton}), we arrive at the multipole expansion
\begin{equation}
\phi(r,\theta,\varphi)
=- \frac{4\pi G}{r}\sum_{\ell m}\frac{1}{2\ell+1}
\int_M \left(\frac{r'}{r}\right)^{\ell}Y_{\ell m}(\theta',\varphi')\rho(\mathbf{x'})\dif\mb{x'}\,Y_{\ell m}(\theta,\varphi), 
\end{equation}
which is valid for $r \ge b$. Differentiating this expression, we see that the normal derivative of the potential can be written as
\begin{equation}
    \left.\frac{\partial \phi}{\partial n}\right|_{\partial B}=\frac{4\pi G}{b^{2}} \sum_{\ell m}\frac{\ell+1}{2\ell+1}\int_M \left(\frac{r'}{b}\right)^{\ell}Y_{\ell m}(\theta',\varphi')\rho(\mathbf{x'})\dif\mb{x'}\,Y_{\ell m}(\theta,\varphi).
\end{equation}
Substituting this expression into eq.(\ref{eq:weak}), we arrive at 
\begin{equation}
\frac{1}{4\pi G}\int_B \nabla\psi\cdot\nabla\phi\dif\mb{x}
= -\int_B \rho \psi\dif\mb{x}
+\sum_{\ell m}
\frac{\ell+1}{(2\ell+1)b^{2}}
\int_{\partial B} Y_{\ell m}\psi\dif S\,
\int_B \left(\frac{r}{b}\right)^{\ell}Y_{\ell m}\rho\dif\mb{x}. 
\label{eq:mp-weak}
\end{equation}
This weak formulation of the problem, which is due to \cite{van2021modelling}, imposes the desired Poisson's equation within $B$ while setting the normal derivative of $\phi$ 
equal to the value provided by the multipole expansion on $\partial B$. 
If we take $\psi$ equal to a constant within this weak formulation and use the orthonormality property of spherical harmonics, we arrive 
at a trivial equality. Thus, the artificial compatibility condition placed on the density within the Neumann version of the domain truncation 
method has been removed. Clearly within numerical work the multipole expansion must be truncated. A special case takes $\ell_{\max} = 0$, 
in which case the reduced form becomes
\begin{equation}
    \frac{1}{4\pi G}\int_M \nabla\psi\cdot\nabla\phi\dif\mb{x}
= -\int_B \rho \psi\dif\mb{x}
+
\frac{1 }{4\pi b^{2}}
\int_{\partial B} \psi\dif S\,
\int_M \rho\dif\mb{x}, 
\end{equation}
which was discussed by \cite{may2011optimal}.

\subsubsection{Numerical implementation}

To implement the multipole method, we consider the bilinear form
\begin{equation}
  (\psi, \rho) \mapsto  \sum_{\ell m}
\frac{\ell+1}{(2\ell+1)b^{2}}
\int_{\partial B} Y_{\ell m}\psi\dif S\,
\int_M \left(\frac{r}{b}\right)^{\ell}Y_{\ell m}\rho\dif\mb{x}, 
\end{equation}
where $\psi$ and $\rho$ are scalar fields, and the truncation of the spherical harmonic summation at some degree, $\ell_{\max}$, is left implicit.
Relative to a finite-element discretisation, this form is associated with a matrix, $\mathbf{B}$, say. This matrix can then act on the 
discretised representation of the density field to produce the necessary modification term to the right hand side within Poisson's equation. 
Equivalently, we could of course choose to directly discretise the linear form 
\begin{equation}
  \psi \mapsto  \sum_{\ell m}
\frac{\ell+1}{(2\ell+1)b^{2}}
\int_{\partial B} Y_{\ell m}\psi\dif S\,
\int_M \left(\frac{r}{b}\right)^{\ell}Y_{\ell m}\rho\dif\mb{x}, 
\end{equation}
for fixed $\rho$. The cost of assembling this linear form or assembling the operator and acting it on a density vector
are broadly similar, but in the latter case there is the advantage that right hand sides for different values of $\rho$
can be readily determined. 

Within the MFEM framework, our implementation of the multipole form  resembles closely the 
approach discussed in detail for the DtN form. As before, we have written  a class that inherits from \texttt{mfem:Operator} 
which implements the action of $\mathbf{B}$ and of its transpose. 
Within the serial version of the class, this matrix is factored as
\begin{equation}
    \mathbf{B} = \mathbf{D}^{T} \mathbf{C}, 
\end{equation}
where $\mathbf{C}$ maps a discretised density field to a vector of its multipole coefficients up to the truncation degree, 
while $\mathbf{D}$ maps a scalar field to its spherical harmonic coefficients on $\partial B$ up to the same degree. 
Note that while $\mathbf{D}$ depends only on degrees of freedom associated with $\partial B$, the action of $\mathbf{C}$
involves all degrees of freedom lying in the interior domain, $M$. To parallelise the action of the discretised multipole form, we 
again proceed in three stages. First, each processor acts its local version of $\mathbf{C}$ on its part of
the discretised density vector. Next,  a call to \texttt{MPI\_Allreduce} is used to sum  the contributions to the multipole 
coefficients across all processors. Finally, the local form of $\mathbf{D}^{T}$  on each 
processor acts on the vector of multipole coefficients to compute its part of the output vector. 

The communication costs and parallel scaling of the multipole operator is very similar to that for 
the DtN method. In the multipole case, assembly of the local matrices, $\mathbf{C}$, 
involves contributions from all processors. Here, it would be possible to define a custom communicator 
that comprises only processors that contain part of the interior domain, $M$, but the practical advantages of 
doing this are less clear. A key difference for static gravity calculations, however, is that only one matrix vector product with the multipole operator is needed to 
update the problem's right hand side. Having done this, the necessary linear system  involves 
just the standard diffusion term occurring within  the Neumann variant of the domain truncation method. 
This means that using the multipole method entails no increase in solver time once the 
necessary right hand side has been constructed. 

Within the libraries FEniCS and Firedrake, we can  provide a workable implementation 
of the multipole method using an operator formed from appropriate tensor products of 
linear forms. In particular, the discretised multipole operator can be written
as the truncated sum
\begin{equation}
    \mathbf{B} = \sum_{\ell m}\mathbf{d}_{\ell m} \otimes \mathbf{c}_{\ell m}, 
\end{equation}
where $\mathbf{c}_{\ell m}$ is the vector corresponding to the linear form that maps 
the density to the $(\ell,m)_{\mathrm{th}}$ multipole coefficient, and $\mathbf{d}_{\ell m}$ 
is the vector corresponding to the linear form that maps a function to 
its spherical harmonic coefficients on $\partial B$. As before, 
the necessary forms can be defined and assembled using the
unified form language of those libraries, while the 
matrix, $\mathbf{B}$, is constructed in an efficient matrix-free 
manner using \texttt{MatCreateLRC} from the PETsc library.

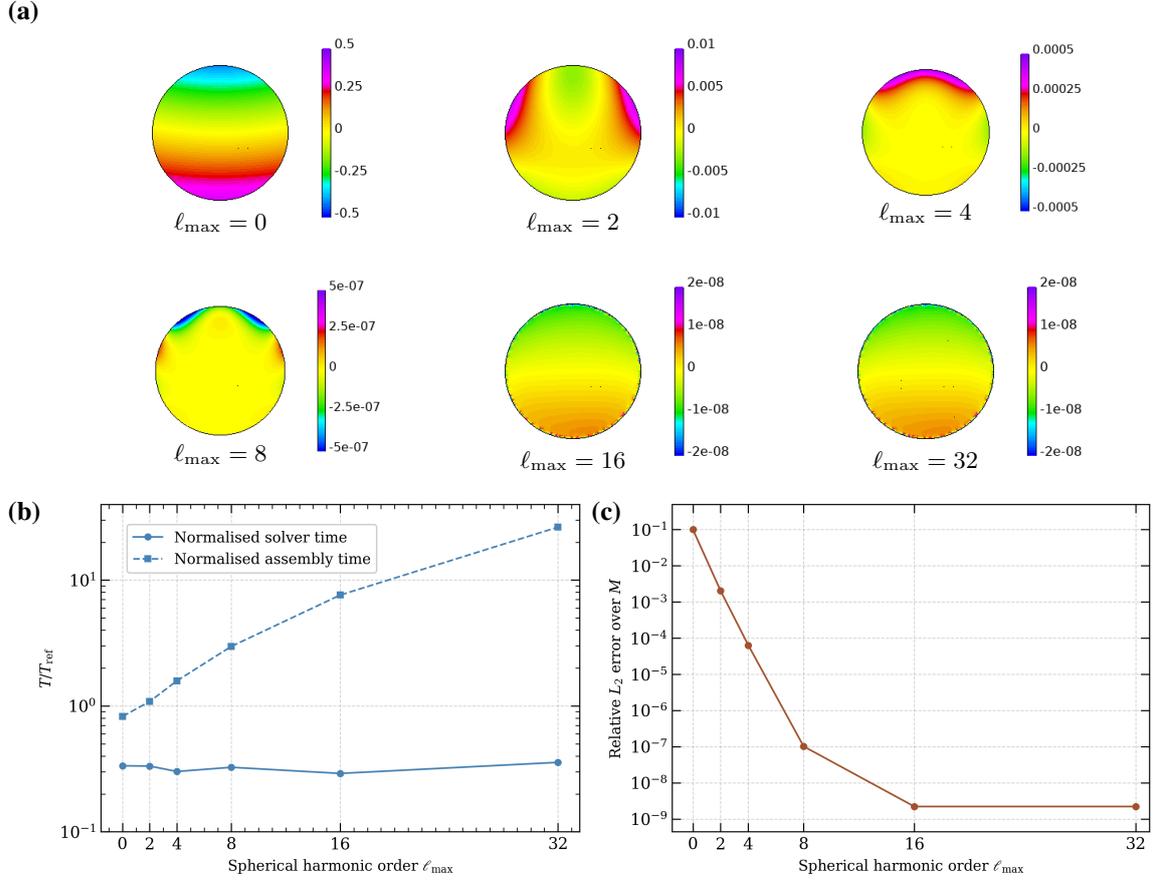
\begin{figure}
\centering

\begin{subfigure}{0.85\textwidth}
  \centering
  \setlength{\tabcolsep}{6pt}
  \renewcommand{\arraystretch}{1.0}

  \begin{tikzpicture}
    \node[anchor=south west, inner sep=0] (grid) at (0,0) {
      \begin{minipage}{\linewidth}
        \centering
        \begin{tabular}{@{}c c c@{}}
          
          \begin{minipage}{0.28\textwidth}\centering
            \begin{tikzpicture}
              \node[anchor=south west, inner sep=0] (p) at (0,0)
                {\includegraphics[width=\linewidth]{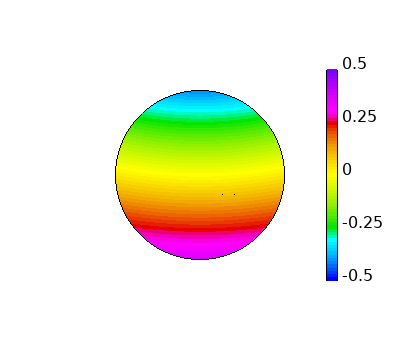}};
            \end{tikzpicture}\\[-4.5em]
            {\normalsize $\ell_{\max}=0$}
          \end{minipage} &
          \begin{minipage}{0.28\textwidth}\centering
            \begin{tikzpicture}
              \node[anchor=south west, inner sep=0] (p) at (0,0)
                {\includegraphics[width=\linewidth]{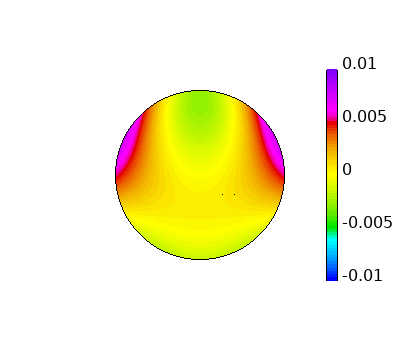}};
            \end{tikzpicture}\\[-4.5em]
            {\normalsize $\ell_{\max}=2$}
          \end{minipage} &
          \begin{minipage}{0.28\textwidth}\centering
            \begin{tikzpicture}
              \node[anchor=south west, inner sep=0] (p) at (0,0)
                {\includegraphics[width=\linewidth]{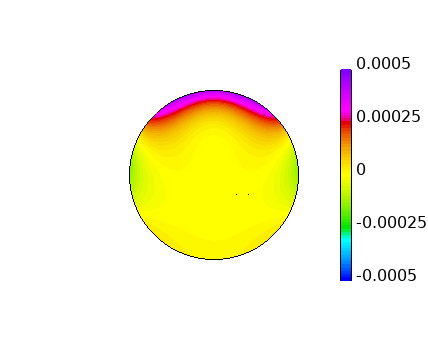}};
            \end{tikzpicture}\\[-4.5em]
            {\normalsize $\ell_{\max}=4$}
          \end{minipage}
          \\
          [0.4em]
          
          \begin{minipage}{0.28\textwidth}\centering
            \begin{tikzpicture}
              \node[anchor=south west, inner sep=0] (p) at (0,0)
                {\includegraphics[width=\linewidth]{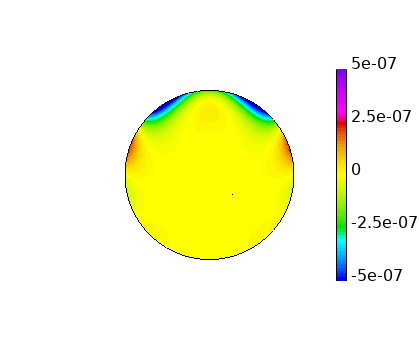}};
            \end{tikzpicture}\\[-4.5em]
            {\normalsize $\ell_{\max}=8$}
          \end{minipage} &
          \begin{minipage}{0.28\textwidth}\centering
            \begin{tikzpicture}
              \node[anchor=south west, inner sep=0] (p) at (0,0)
                {\includegraphics[width=\linewidth]{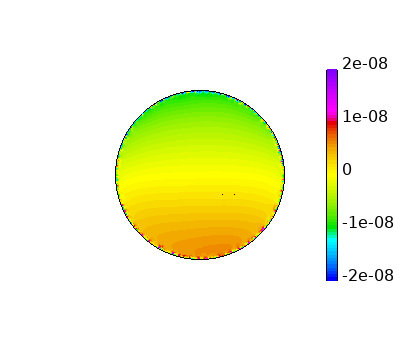}};
            \end{tikzpicture}\\[-4.5em]
            {\normalsize $\ell_{\max}=16$}
          \end{minipage} &
          \begin{minipage}{0.28\textwidth}\centering
            \begin{tikzpicture}
              \node[anchor=south west, inner sep=0] (p) at (0,0)
                {\includegraphics[width=\linewidth]{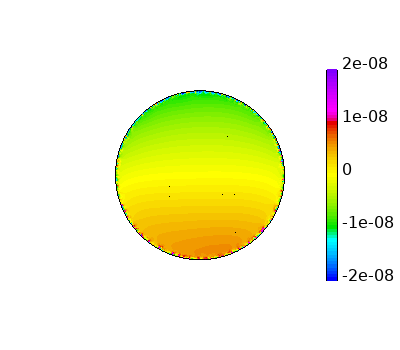}};
            \end{tikzpicture}\\[-4.5em]
            {\normalsize $\ell_{\max}=32$}
          \end{minipage}
        \end{tabular}
      \end{minipage}
    };
    
    \begin{scope}[x={(grid.south east)}, y={(grid.north west)}]
      \node[anchor=north west, font=\large\bfseries] at (0,1) {(a)};
    \end{scope}
  \end{tikzpicture}
\end{subfigure}

\vspace{1.0em}

\begin{subfigure}{0.85\textwidth}
  \centering
  \begin{minipage}{0.49\textwidth}
    \centering
    \begin{tikzpicture}
      \node[anchor=south west, inner sep=0] (imgb) at (0,0)
        {\includegraphics[width=\linewidth]{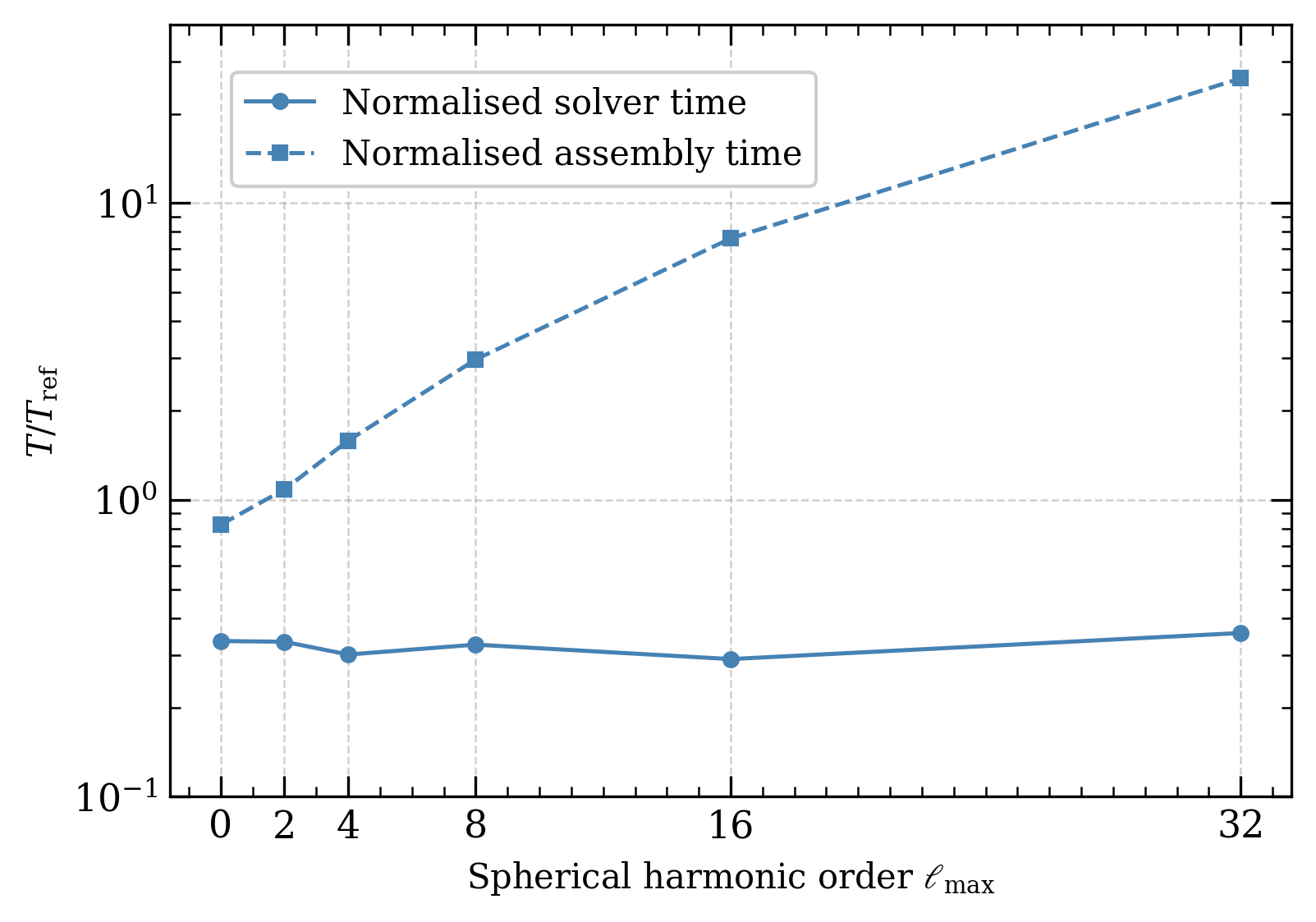}};
      \begin{scope}[x={(imgb.south east)}, y={(imgb.north west)}]
        \node[anchor=north west, font=\large\bfseries] at (-0.05,1) {(b)};
      \end{scope}
    \end{tikzpicture}
  \end{minipage}\hfill
  \begin{minipage}{0.49\textwidth}
    \centering
    \begin{tikzpicture}
      \node[anchor=south west, inner sep=0] (imgc) at (0,0)
        {\includegraphics[width=\linewidth]{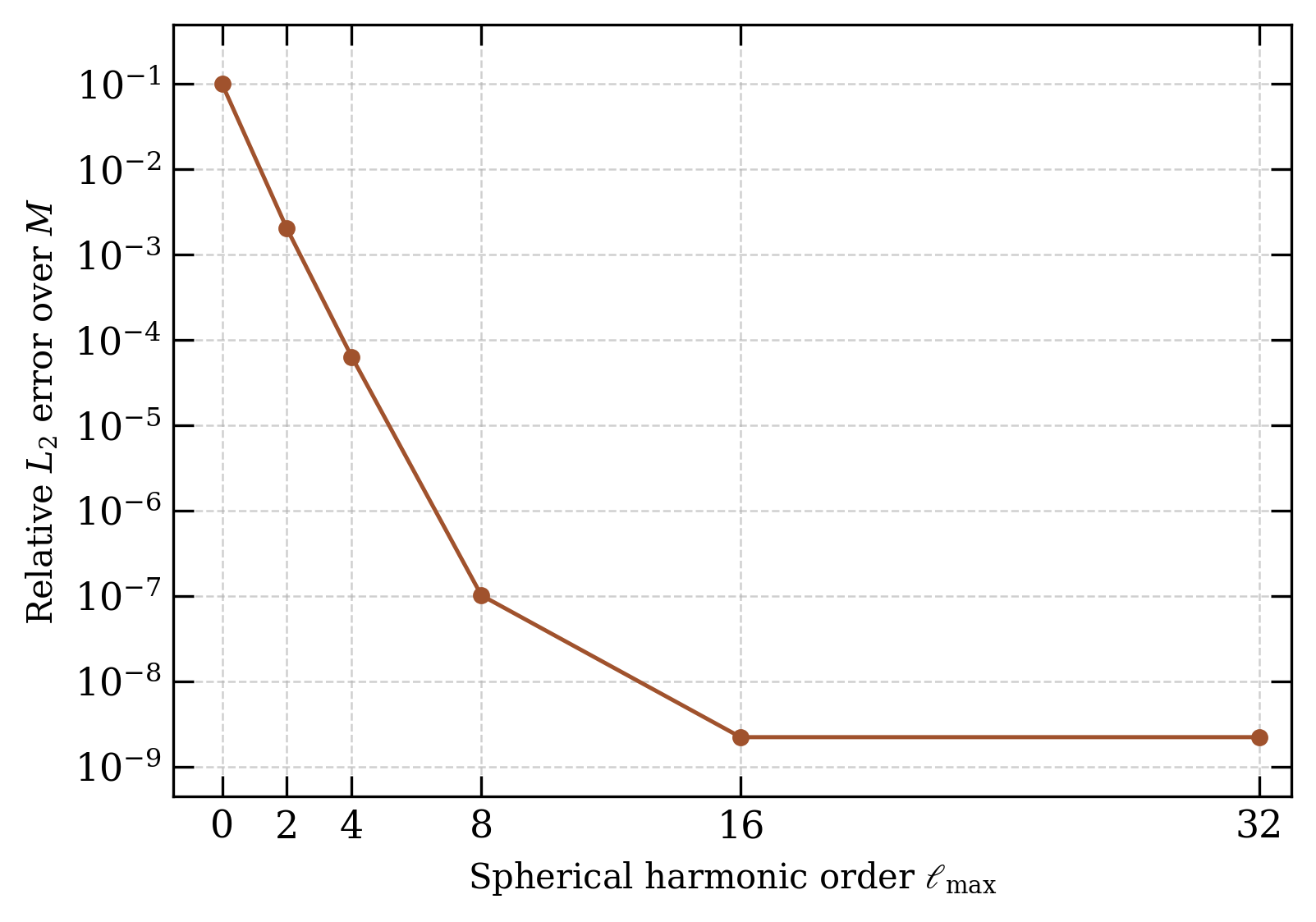}};
      \begin{scope}[x={(imgc.south east)}, y={(imgc.north west)}]
        \node[anchor=north west, font=\large\bfseries] at (-0.025,1) {(c)};
      \end{scope}
    \end{tikzpicture}
  \end{minipage}
\end{subfigure}

\caption{Multipole expansion method for the offset-sphere configuration with ${a/b=10/7}$. (a) Signed relative error distributions over $M$ for six spherical harmonic orders (\(\ell_{\max}=\{0,2,4,8,16,32\}\)). (b) Assembly time (normalised by the time for assembling the diffusion term with the zero Dirichlet condition, $T_{ref}=3.81\mathrm{s}$) and solver time (normalised by that achieving $\varepsilon_{L^2(M)}=1.06\times 10^{-6}$ with a large domain of $b/a=50$, $T_{ref}=56.01\mathrm{s}$) versus \(\ell_{\max}\). All times are averaged over 20 runs with 8 CPU processors. (c) Relative \(L_2\) error over $M$, $\varepsilon_{L^2(M)}$, versus \(\ell_{\max}\).}
\label{fig:multipole}
\end{figure}

\subsubsection{Benchmarks}
Fig.\ref{fig:multipole} illustrates the performance of the multipole method for the offset-sphere benchmark. These calculations 
were performed using the same mesh  as for the DtN tests, and the calculations again run on 8 CPUs. As for DtN method, the relative error decreases rapidly
as $\ell_{\mathrm{max}}$ increases. Moreover, for the same truncation degree, the relative error for the DtN and multipole methods are very close. From Fig.~\ref{fig:multipole}(b), the solver time for the multipole method is nearly constant, consistent with the fact that the stiffness matrix remains unchanged; this time is always below 40\% of that required to achieve $10^{-6}$ accuracy using large-domain truncation, let alone the $10^{-9}$ accuracy attainable with the multipole approach. Nevertheless, the assembly time increases rapidly with $\ell_{\mathrm{max}}$ and, at $\ell_{\mathrm{max}} = 32$, exceeds the stiffness-matrix assembly time by a factor of twenty.

Most of the general considerations discussed in the context of the DtN method apply equally to the multipole case. In particular, the need to include higher harmonic 
degrees for accuracy can be mitigated by increasing the radius of the computational domain. The result is that both the DtN and multipole methods provide accurate and 
efficient approaches for large-scale static gravity calculations.

\subsection{Linearised problem}\label{sec:lin}
\subsubsection{Theory}

We now consider the linearised form of Poisson's equation associated with the application of a displacement field, $\mathbf{u}$, to a planet, $M$, with equilibrium density
$\rho$. We now write $\phi^{0}$ for the equilibrium potential associated with $\rho$, and  let $\phi$ denote the  first order perturbation to the gravitational potential
caused by the displacement. It is well known \citep[e.g.][]{dahlen1999theoretical} the perturbed field satisfies the Poisson equation
\begin{equation}
    \Delta \phi = -4\pi G \nabla\cdot(\rho \mathbf{u}), 
\end{equation}
while across any material discontinuities (such as $\partial M$) we have the jump conditions
\begin{equation}
\left[\phi\right]_{-}^{+} = 0, \quad \left[\hat{\mathbf{n}}\cdot \nabla \phi + 4\pi G \rho \hat{\mathbf{n}} \cdot \mathbf{u}
\right]_{-}^{+}   = 0,
\end{equation}
where $\hat{\mathbf{n}}$ is the outward unit normal on the boundary; note that by convention, we extend $\rho$ to be zero outside of $M$.

As before, let $B$ be a bounded computational domain that contains $M$. For a suitably regular test function, $\psi$, 
then through simple integrations by parts we arrive at 
    \begin{equation}
\frac{1}{4\pi G}\int_{B}\nabla\psi\cdot\nabla\phi\dif\mb{x} -\frac{1}{4\pi G}\int_{\partial B}\psi\frac{\partial \phi}{\partial n}\dif S = 
-\int_M \rho  \nabla \psi \cdot \mathbf{u}\dif\mb{x}, 
\label{eq:weak_lin}
\end{equation}
which provides the starting point for various weak formulations of the problem. Comparison of eq.(\ref{eq:weak}) and (\ref{eq:weak_lin})
shows that the only difference between the static and linearised Poisson's equations lies within the force term.

\subsubsection{The domain truncation method}

Either the Dirichlet or Neumann variants of the domain truncation method can provide a simple, 
though unless the computational domain is very large, usually rather inaccurate solution of the linearised Poisson equation. In both cases, the 
weak form is reduced to 
\begin{equation}
    \frac{1}{4\pi G}\int_{B}\nabla\psi\cdot\nabla\phi\dif\mb{x}  = 
-\int_M \rho  \nabla \psi \cdot \mathbf{u}\dif\mb{x}, 
\end{equation}
where in the Dirichlet case the same homogeneous boundary conditions are imposed on the test functions. 
It is notable that within the Neumann case there is no longer a compatibility issue that must 
be considered. Indeed, if we take $\psi$ a constant in the above weak form, we arrive at
a trivial equality. Thus either the Dirichlet or Neumann variants of the domain truncation 
method can be applied to the linearised problem for any choice of displacement field. Indeed, the Neumann method is 
generally preferable because the normal derivative of the exterior potential decreases 
more rapidly than the potential itself. 

\subsubsection{The Dirichlet-to-Neumann method}

Applying the Dirichlet to Neumann mapping to specify the normal derivative in eq.(\ref{eq:weak_lin}) we arrive at
\begin{equation}
    \frac{1}{4\pi G} \int_B\nabla \psi\cdot\nabla \phi \dif \mathbf{x} + \frac{1}{4\pi G}\sum_{\ell m}(\ell+1)b\,\psi_{\ell m}(b)\phi_{\ell m}(b) =
    -\int_M \rho  \nabla \psi \cdot \mathbf{u}\dif\mb{x}.
\end{equation}
The bilinear forms on the left hand side are identical to those within the DtN version of the static problem, and 
hence the methods discussed for that case carry over without essential modification.

\subsubsection{The Multipole method}

The multipole method is based on a solution to the exterior problem expressed as a multipole expansion. The form 
of this solution changes within the linearised problem, and hence this method discussed for the 
static problem does require modification. In terms of the fundamental solution, $\Phi$, defined previously, 
the solution to the linearised problem can be written as
\begin{equation}
     \phi(\mathbf{x}) = -4\pi G\int_M \Phi(\mathbf{x},\mathbf{x'})\,\nabla'\cdot(\rho'\mathbf{u}')\dif\mathbf{x'}
     + 4\pi G \int_{\partial M} \Phi(\mathbf{x},\mathbf{x'})\left[\rho'\hat{\mathbf{n}}'\cdot\mathbf{u}'\right]_{-}^{+}\dif S',
\end{equation}
where primes are used to denote functions that depend on the dummy integration variable, and $\nabla'$ is a derivative
taken with respect to this variable \citep[e.g.][]{dahlen1999theoretical}. Note that if the model possesses additional 
material discontinuities across which the density is discontinuous, then these must be similarly 
included. Performing an integration by parts, the above expression reduces to 
\begin{equation}
    \phi(\mathbf{x}) = 4\pi G \int_{M}  \nabla' \Phi(\mathbf{x},\mathbf{x}') \cdot  \rho' \mathbf{u}' \dif\mathbf{x}'.
\end{equation}
Within this latter expression, the contribution from $\partial M$ or from any internal boundaries, 
have been incorporated implicitly. Using the multipole expansion of the fundamental solution in 
eq.(\ref{eq:fund_sh}), we can then write the normal derivative of the 
exterior potential as a sum over multipole coefficients defined in terms of the displacement vector, $\mathbf{u}$.
This result can be conveniently written as 
\begin{equation}
    \frac{1}{4\pi G}\int_{\partial B}\psi\frac{\partial \phi}{\partial n}\dif S = \sum_{\ell m}\int_{\partial B}Y_{\ell m}\psi\dif S\,
    \frac{\ell+1}{2\ell+1}\frac{1}{b^3}\int_M\left(\frac{r}{b}\right)^{\ell-1}\left(\ell Y_{\ell m}\hat{\mathbf{r}}+\nabla_1Y_{\ell m}\right)\cdot\mathbf{u}\dif\mathbf{x},
\end{equation}
for any test function $\psi$, where $\hat{\mathbf{r}}$ is the radial unit vector, and $\nabla_{1}$ denotes the tangential gradient operator on the unit sphere.  
Using this expression within eq.(\ref{eq:weak_lin}), we arrive at a weak formulation of Poisson's equation that accounts for the 
exterior domain exactly. The application of the multipole method to the linearised Poisson equation was, to our knowledge, first discussed by 
\cite{van2021modelling}. Within that paper, however, the contribution to the exterior solution associated with the jump in the normal derivative 
of $\phi$ across material discontinuities was neglected (see their equations 4 and 5). It is not clear whether this omission is
just an oversight in the paper or if it reflects a deficiency within their numerical implementation of the method. 

\subsubsection{Numerical implementation}

It is only the numerical implementation of the multipole method within the linearised problem that requires discussion. Here, 
we need to implement the bilinear form
\begin{equation}
    (\psi,\mathbf{u}) \mapsto  \sum_{\ell m}\int_{\partial B}Y_{\ell m}\psi\dif S\,
    \frac{\ell+1}{2\ell+1}\frac{1}{b^3}\int_M\left(\frac{r}{b}\right)^{\ell-1}\left(\ell Y_{\ell m}\hat{\mathbf{r}}+\nabla_1Y_{\ell m}\right)\cdot\mathbf{u}\dif\mathbf{x}
    \label{eq:mult_lin}
\end{equation}
with the summation truncated at some degree, $\ell_{\max}$. This is done using the same approach as for the multipole form within the static gravity problem.
The parallel scaling of the method is unchanged, but the number of floating point operations is increased due to the need 
to account for vector finite-element spaces. During the assembly of the necessary matrices, we again generate associated 
Legendre polynomials using stable upward recursion in $\ell$,  while their co-latitudinal derivatives
are found using eq.(C.119) of \cite{dahlen1999theoretical}.

As with the multipole method for static problems, it would be possible to  assemble the linear form corresponding to eq.(\ref{eq:mult_lin})
for a fixed value of the displacement vector. In practice, however, building an operator that acts on the discretisation of $\mathbf{u}$
is of greater utility. This is because the main application of the linearised Poisson problem is  to  coupled elasticity (or viscoelasticity) with self-gravitation,
and here repeated actions of this operator are required within the solution of the coupled linear equations. 

\begin{figure}
\centering
\begin{subfigure}{0.85\textwidth}
  \centering
  \setlength{\tabcolsep}{6pt}
  \renewcommand{\arraystretch}{1.0}

  \begin{tikzpicture}
    \node[anchor=south west, inner sep=0] (grid) at (0,0) {
      \begin{minipage}{\linewidth}
        \centering
        \begin{tabular}{@{}c c c@{}}
          
          \begin{minipage}{0.28\textwidth}\centering
            \includegraphics[width=\linewidth]{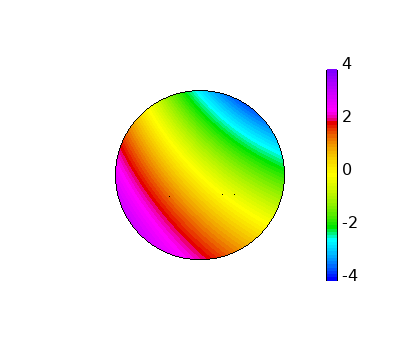}\\[-4.5em]
            {\normalsize $\ell_{\max}=0$}\\
          \end{minipage} &
          \begin{minipage}{0.28\textwidth}\centering
            \includegraphics[width=\linewidth]{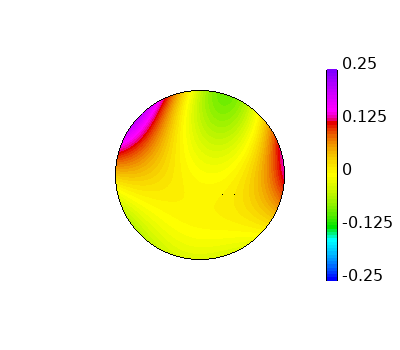}\\[-4.5em]
            {\normalsize $\ell_{\max}=2$}\\
          \end{minipage} &
          \begin{minipage}{0.28\textwidth}\centering
            \includegraphics[width=\linewidth]{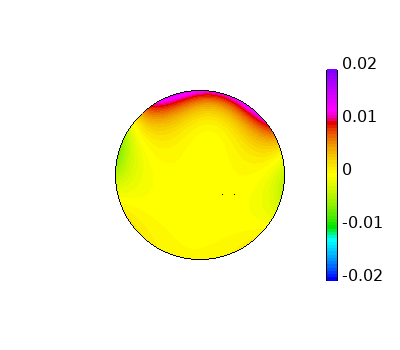}\\[-4.5em]
            {\normalsize $\ell_{\max}=4$}\\
          \end{minipage}
          \\
          [0.4em]
          
          \begin{minipage}{0.28\textwidth}\centering
            \includegraphics[width=\linewidth]{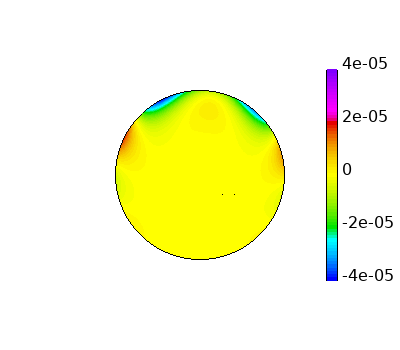}\\[-4.5em]
            {\normalsize $\ell_{\max}=8$}\\
          \end{minipage} &
          \begin{minipage}{0.28\textwidth}\centering
            \includegraphics[width=\linewidth]{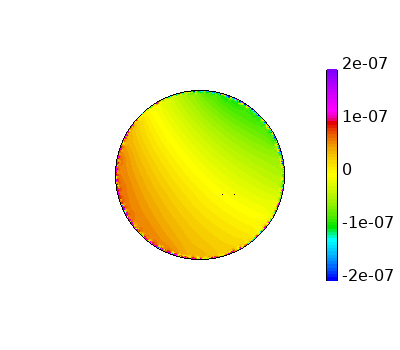}\\[-4.5em]
            {\normalsize $\ell_{\max}=16$}\\
          \end{minipage} &
          \begin{minipage}{0.28\textwidth}\centering
            \includegraphics[width=\linewidth]{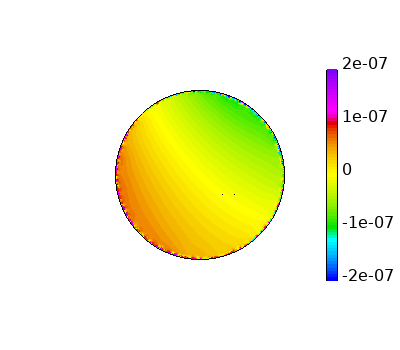}\\[-4.5em]
            {\normalsize $\ell_{\max}=32$}\\
          \end{minipage}
        \end{tabular}
      \end{minipage}
    };
    \begin{scope}[x={(grid.south east)}, y={(grid.north west)}]
      \node[anchor=north west, font=\large\bfseries] at (0,1) {(a)};
    \end{scope}
  \end{tikzpicture}
\end{subfigure}

\vspace{1.0em}

\begin{subfigure}{0.85\textwidth}
  \centering
  \begin{minipage}{0.49\textwidth}
    \centering
    \begin{tikzpicture}
      \node[anchor=south west, inner sep=0] (imgb) at (0,0)
        {\includegraphics[width=\linewidth]{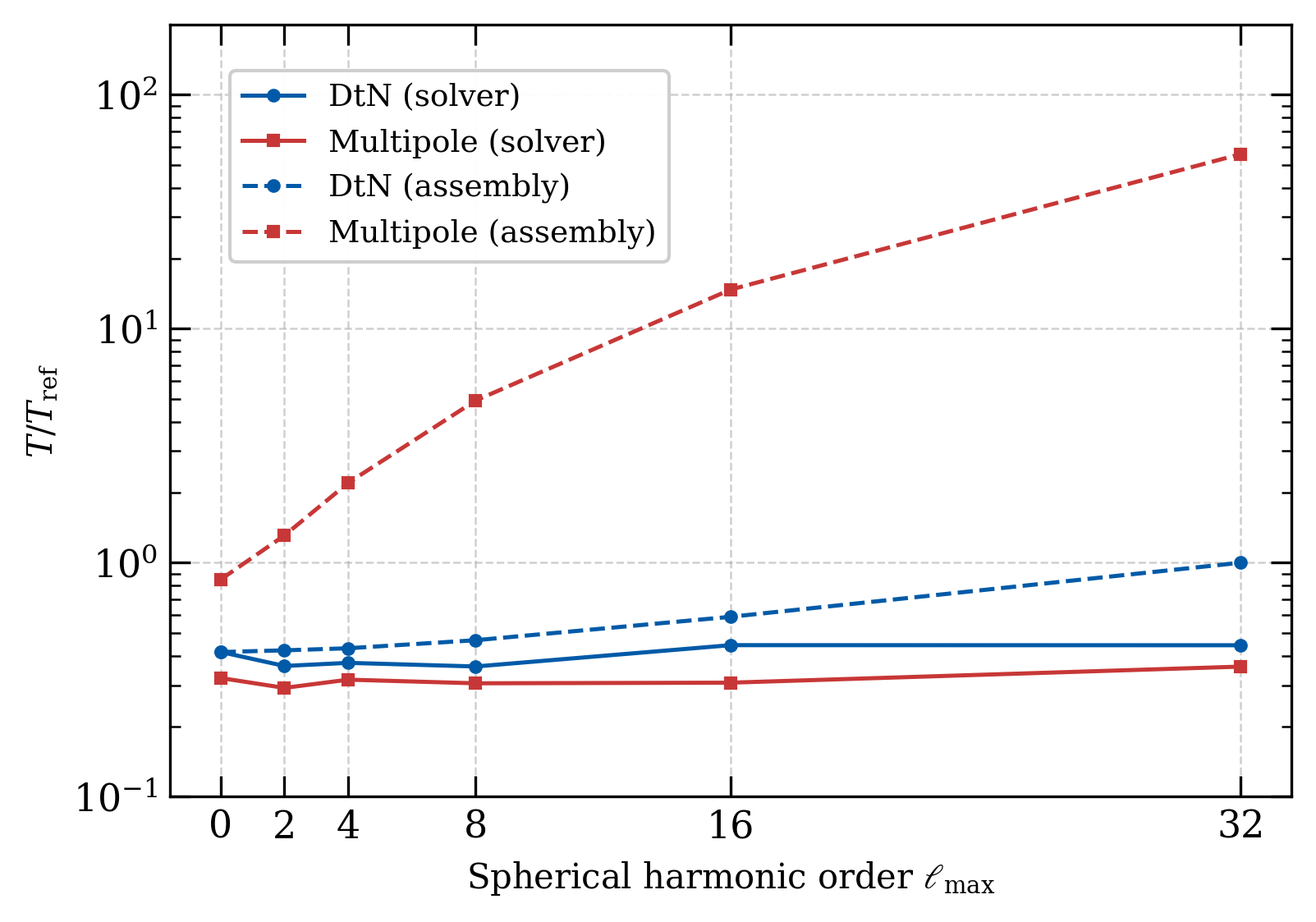}};
      \begin{scope}[x={(imgb.south east)}, y={(imgb.north west)}]
        \node[anchor=north west, font=\large\bfseries] at (0,1) {(b)};
      \end{scope}
    \end{tikzpicture}
  \end{minipage}\hfill
  \begin{minipage}{0.49\textwidth}
    \centering
    \begin{tikzpicture}
      \node[anchor=south west, inner sep=0] (imgc) at (0,0)
        {\includegraphics[width=\linewidth]{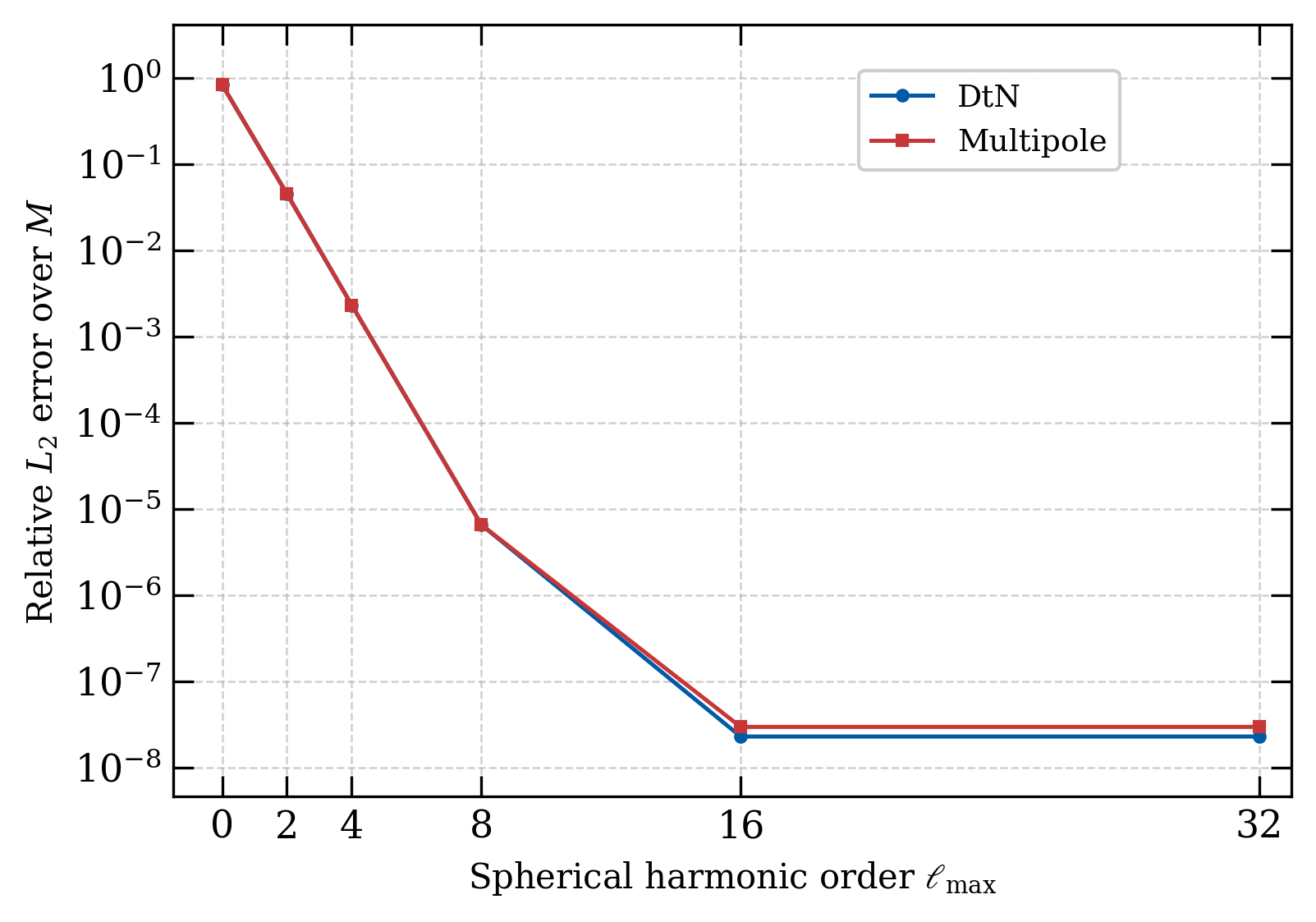}};
      \begin{scope}[x={(imgc.south east)}, y={(imgc.north west)}]
        \node[anchor=north west, font=\large\bfseries] at (0,1) {(c)};
      \end{scope}
    \end{tikzpicture}
  \end{minipage}
\end{subfigure}

\caption{Linearised gravity problem on the offset-sphere configuration with presumed deformation $\mathbf{u}=(1,1,1)b$.
(a) Signed relative error distributions over the offset sphere $M$ for the multipole method at six spherical harmonic orders (\(\ell_{\max}=\{0,2,4,8,16,32\}\)).
(b) Assembly time (dashed, normalised by that for assembling the diffusion term with the zero Dirichlet condition, $T_{ref}=3.81\mathrm{s}$) and solver time (normalised by that achieving $\varepsilon_{L^2(M)}=1.06\times 10^{-6}$ with a large domain of $b/a=50$, $T_{ref}=56.01\mathrm{s}$) versus \(\ell_{\max}\) comparing the DtN (blue) and multipole (red) methods.
(c) Relative \(L_2\) error over $M$ versus \(\ell_{\max}\) for both methods.}
\label{fig:linearised_compare}
\end{figure}

\subsubsection{Benchmarks}

To set up a simple benchmark for numerical solutions of the linearised problem, we use the fact that if $\phi^{0}$ is a solution of 
the static problem with density $\rho$ within $M$, then the solution to the linearised problem for a \emph{constant} displacement field, 
$\mathbf{u}$, takes the simple form
\begin{equation}
    \phi = -\mathbf{u} \cdot \nabla \phi^{0}.
\end{equation}
Applying this idea to the offset-sphere problem considered earlier, our reference solution 
for the linearised problem is then given by
\begin{equation}\label{eq:ana-lin}
\phi_{\text{lin}}(\mathbf{x}) =
\begin{cases}
-\dfrac{4\pi G \rho_{0}}{3}\, (\mathbf{u}\!\cdot\!\hat{\mathbf{r}})\, r, & r \le a, \\[8pt]
-\dfrac{4\pi G \rho_{0} a^{3}}{3}\, \dfrac{\mathbf{u}\!\cdot\!\hat{\mathbf{r}}}{r^{2}}, & r > a,
\end{cases}
\end{equation}
where $r$ is the radial distance from the centre of the sphere $M$ with radius $a$, and
$\hat{\mathbf{r}}$ is the corresponding radial unit vector. In these examples, 
we took $\mathbf{u} = (1, 1, 1)b$, but this choice is immaterial. 
Calculations were done using the same mesh as for the static benchmarks, using 8 CPUs. Within Fig.\ref{fig:linearised_compare} 
results are shown for the DtN and multipole methods, with the comparative performance being 
very similar to that seen within the static benchmarks. Indeed, the main difference is an 
expected increase in the assembly time for the multipole operator due to the need 
for additional floating point operations per element.  

\begin{figure}
\centering
\includegraphics[width=0.5\linewidth]{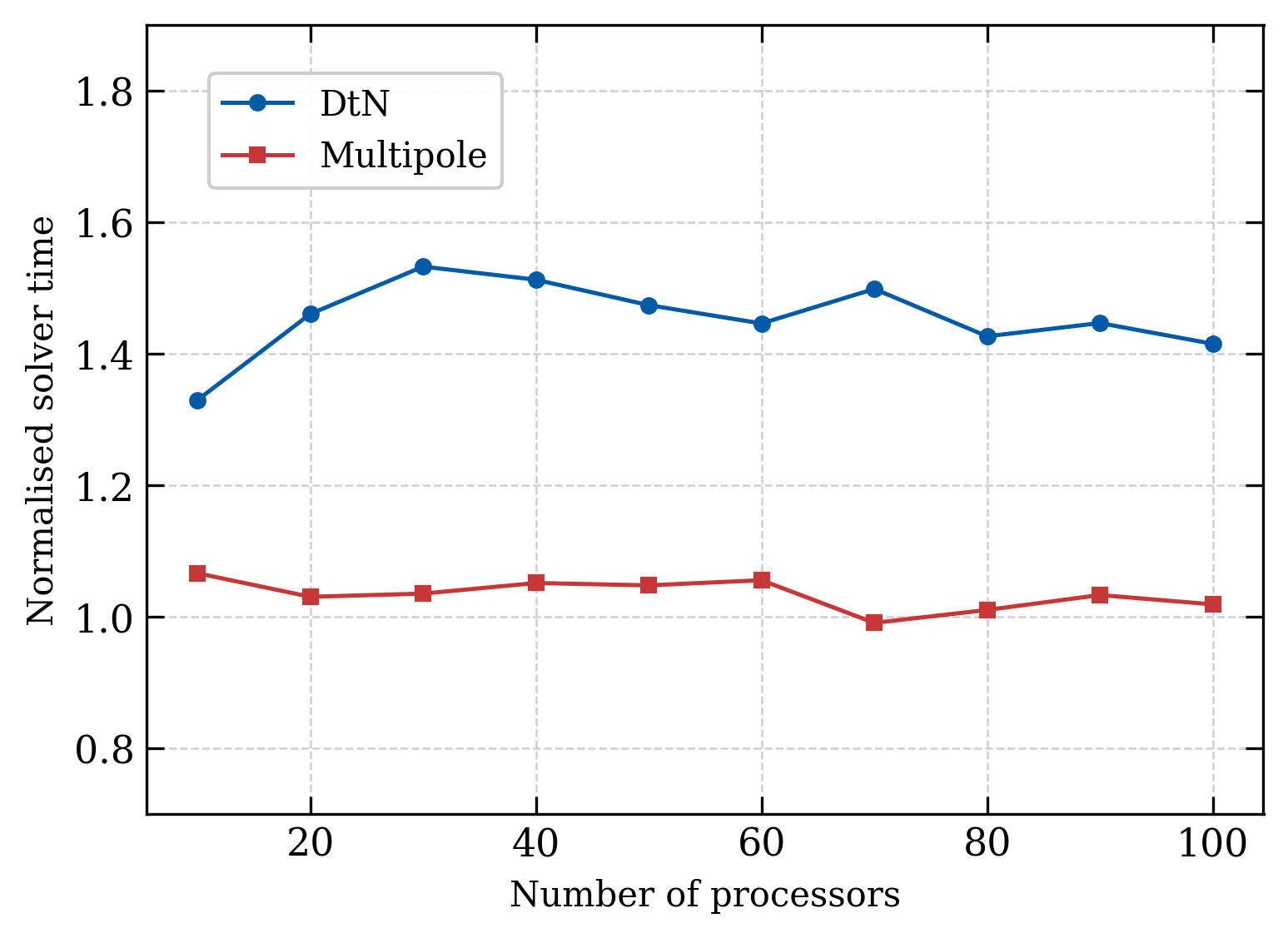}
\caption{Solver time (normalised by that with the zero Dirichlet condition under the same number of processors) versus number of processors for DtN (blue) and multipole (red) methods. Each time is averaged over 20 runs.}
\label{fig:ws}
\end{figure}

\subsection{Parallel scaling}\label{sec:par}

All calculations shown so far have used only 8 CPUs. We noted, however, that both the DtN and multipole methods 
involve non-local communication that has the potential to degrade the parallel performance as the number of
processors is increased. To address this point, we have tested the strong scaling behaviour of the two methods, 
solving the same problem using progressively more CPUs. These results are summarised within Fig.\ref{fig:ws}, 
where we show solver times for the DtN and multipole methods applied to our static benchmark problem
using $\ell_{\max} = 32$. These
solver times are shown relative to the time for solving a simple Poisson problem with homogeneous Dirichlet 
conditions on the same mesh. For the multipole method,  the relative solver time is close to one, this being 
expected as there is no need for non-local  communication during the iterative solution of the equations. 
More significantly, for the range of CPU numbers considered, there is no systematic increase in
the relative solver time for the DtN method. This suggests for the problems considered the communication
between boundary processors necessary within the DtN method is not a limiting factor. Rather, 
the near constant increase in solver time within the DtN method seems mostly to reflect the additional floating point 
operations required. It remains to be verified whether the performance of the DtN method is still competitive 
for very large-scale problems that require thousands of CPU cores. But the results shown are already
sufficient  to demonstrate that the DtN method is suitable within realistic geophysical calculations, with this point emphasised further by the examples discussed in the next section.

\section{Further examples}
\label{sec:applications}

Within this section, we illustrate the potential of the DtN method by using it within two more realistic geophysical applications. 
All calculations have been performed on 8 CPU cores, but the methods can readily be extended to larger scale parallel calculations.

\subsection{Spherically symmetric gravity within PREM}\label{sec:prem}

\begin{figure}
\centering
\raisebox{21pt}{
\begin{subfigure}[b]{0.47\textwidth}
  \centering
  \setlength{\unitlength}{1pt}
  \includegraphics[width=\linewidth]{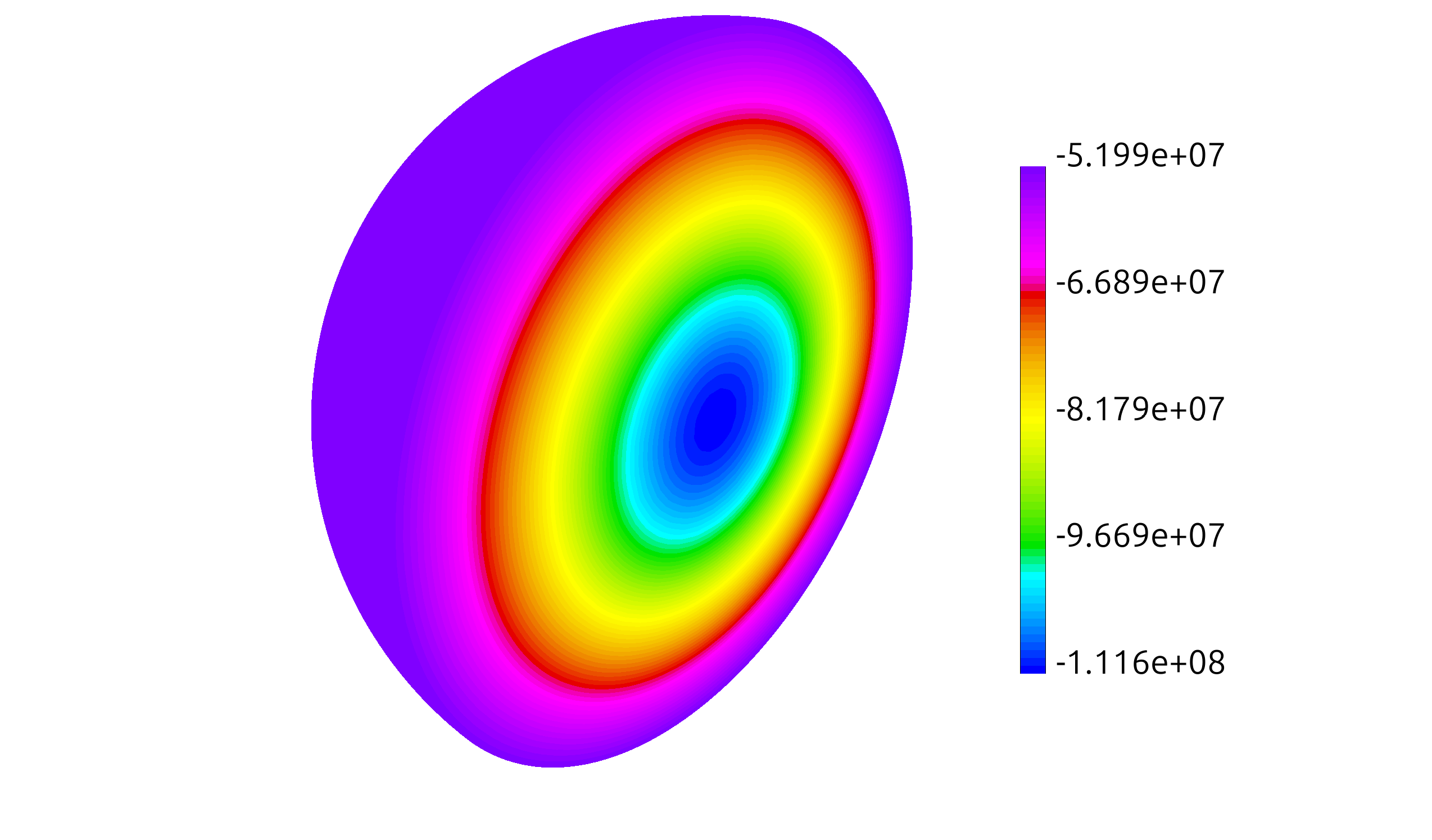}
  \begin{picture}(0,0)
    \put(-100,177){\large\textbf{(a)}}
    \put(44,140){\scriptsize$\mathrm{m}^2/\mathrm{s}^2$}
  \end{picture}
\end{subfigure}\hfill
}
\begin{subfigure}[b]{0.47\textwidth}
  \centering
  \setlength{\unitlength}{1pt}
  \includegraphics[width=\linewidth]{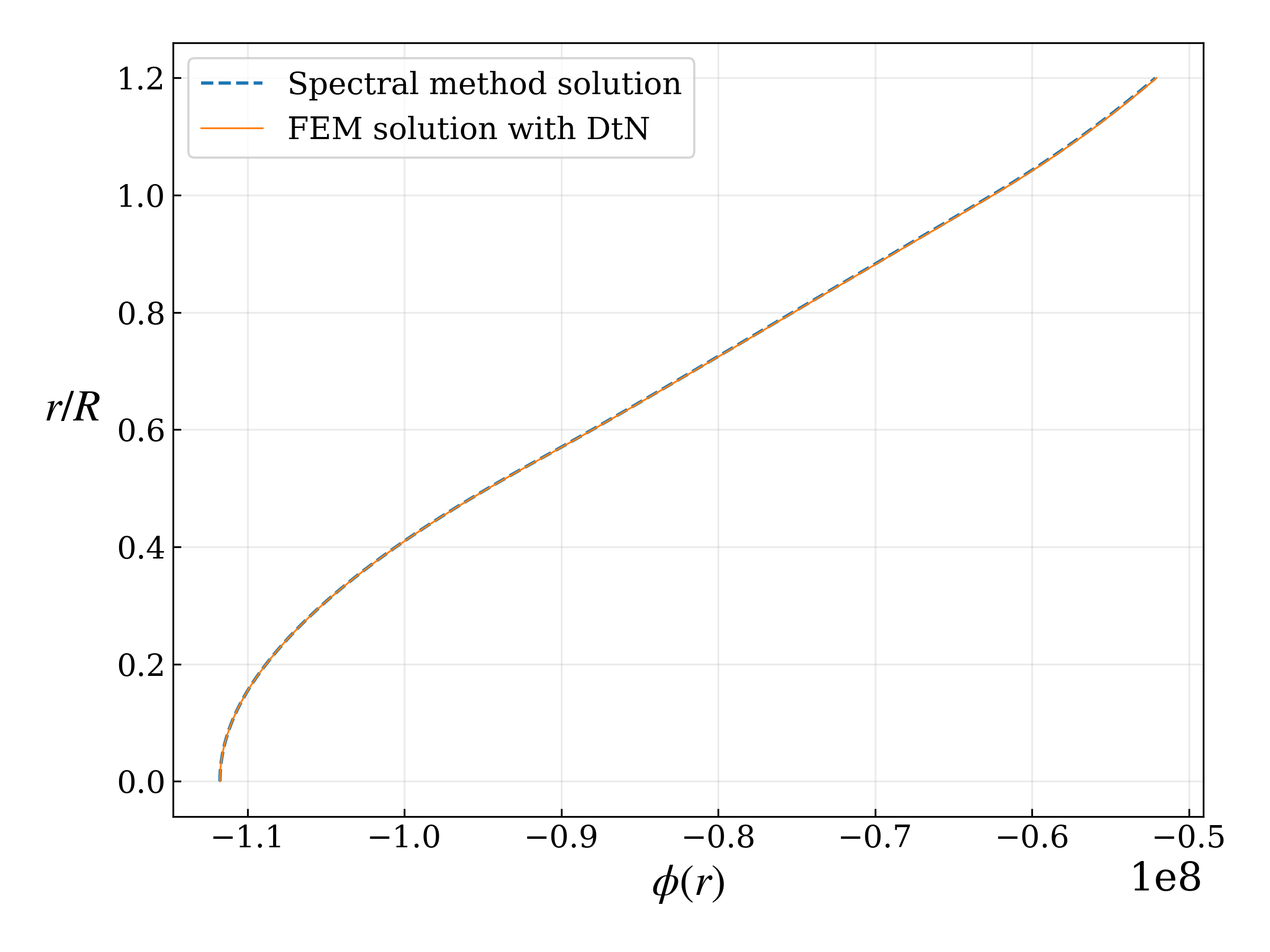}
  \begin{picture}(0,0)
    \put(-125,197){\large\textbf{(b)}}
  \end{picture}
\end{subfigure}
\vspace{-21pt}
\caption{(a) Representative FEM solution on the PREM Earth model. The colour indicates the magnitude of gravity potential in SI units ($\mathrm{m^2/s^2}$). 
(b) Comparison of the Radial profiles of gravity potential $\phi(r)$ between the finite-element solution via the DtN method (solid) and the spectral-method solution (dashed) as in \cite{myhill2025}. The length scale $R=6371\mathrm{km}$.}
\label{fig:prem}
\end{figure}
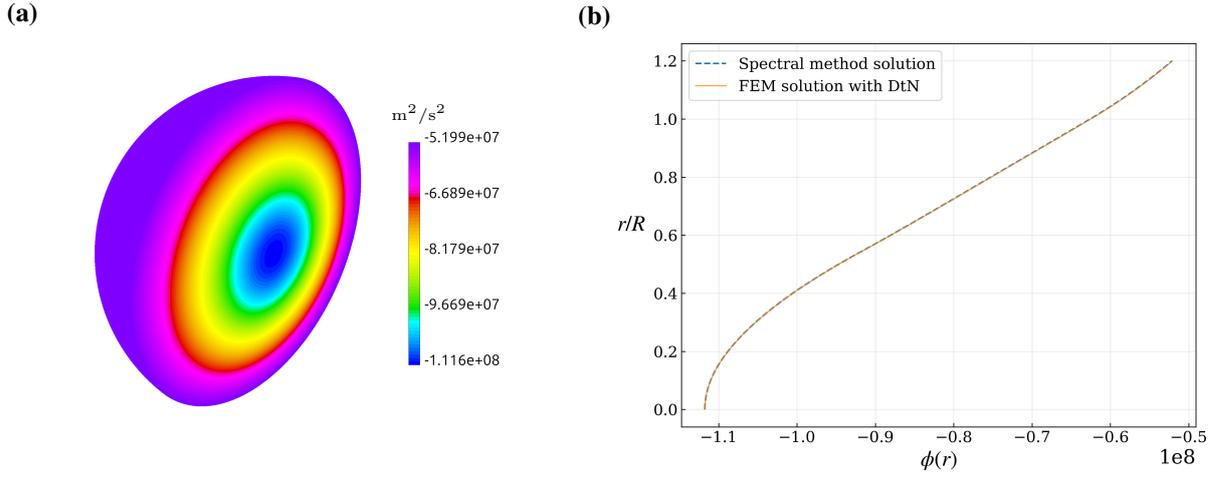

 In the first example, we use the DtN method to 
 determine the static gravitational potential for PREM \citep{dziewonski1981preliminary}. To do this, we constructed a 3D mesh of PREM using gmsh that 
 honours all internal discontinuities. The mesh also includes a spherical buffer layer with outer radius $b=1.2a$, with $a$ the Earth's mean radius, where 
the DtN operator is defined. To assess the accuracy of the DtN solution, we compare a radial profile of the gravity potential to that obtained using a simple radial finite-element method as described in \cite{myhill2025}. As shown in Fig.\ref{fig:prem}(b), the two solutions agree very closely, this providing an independent test of our new finite-element implementations.

\subsection{Phobos}\label{sec:phobos}

\begin{figure}
\centering

\begin{minipage}[t]{0.48\textwidth}
\centering
\raisebox{21pt}{
\begin{tikzpicture}
  \node[anchor=south west, inner sep=0] (A) at (0,0)
  {\includegraphics[
    width=0.9\linewidth,
    trim=0cm 0cm 0cm 0cm,
    clip
  ]{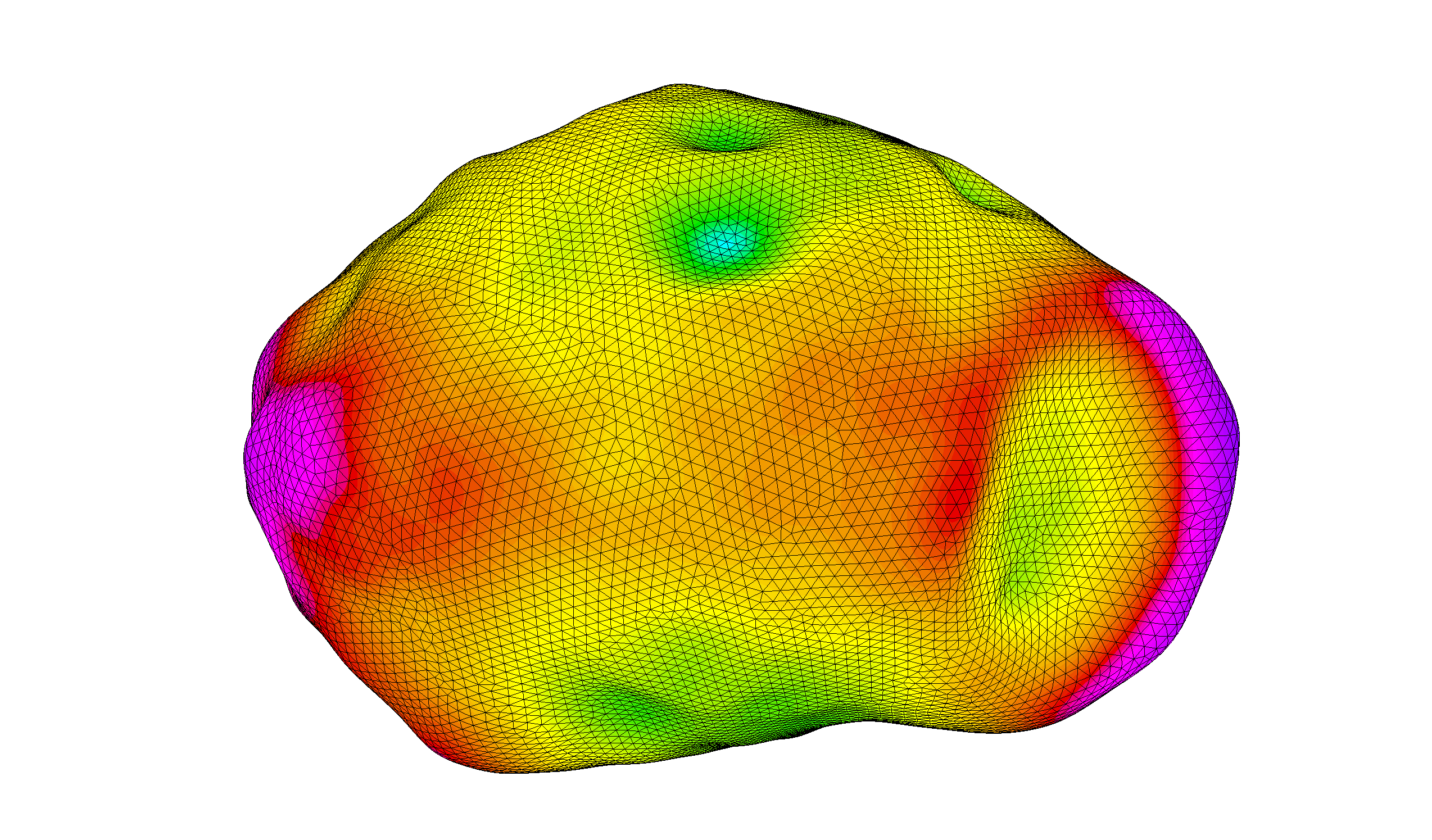}};
  \node[anchor=south west] at (7,0.4)
  {\includegraphics[height=0.4\linewidth]{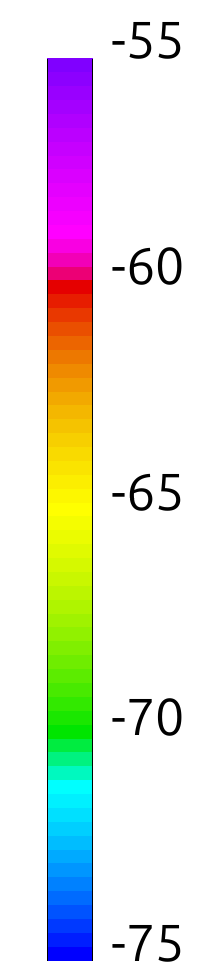}};
  \put(198,115){\small$\mathrm{m}^2/\mathrm{s}^2$}
  \begin{scope}[x={(A.south east)}, y={(A.north west)}]
    \node[anchor=north west, font=\large\bfseries] at (0.03,1.0) {(a)};
  \end{scope}
\end{tikzpicture}
}
\end{minipage}
\hfill
\begin{minipage}[t]{0.48\textwidth}
\centering
\begin{tikzpicture}
  \node[anchor=south west, inner sep=0] (B) at (0,0)
  {\includegraphics[
    width=\linewidth,
    trim=0cm 0cm 0cm 0cm,
    clip
  ]{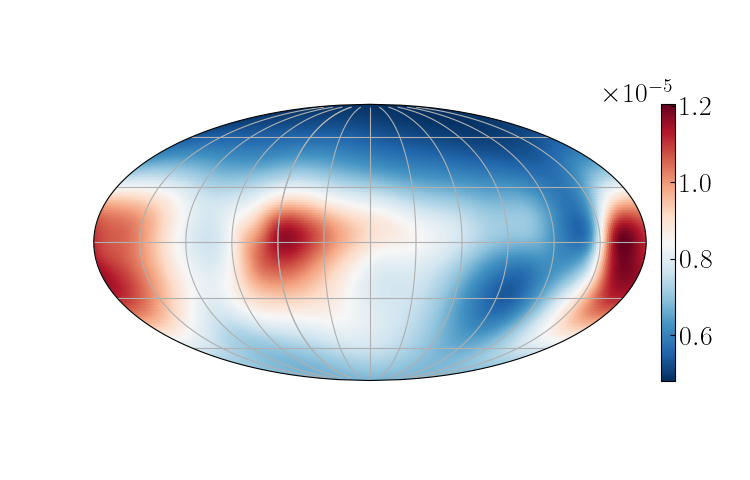}};
  \begin{scope}[x={(B.south east)}, y={(B.north west)}]
    \node[anchor=north west, font=\large\bfseries] at (0.03,0.94) {(b)};
  \end{scope}
\end{tikzpicture}
\end{minipage}

\caption{(a) Mesh and computed gravity potential over the surface of Phobos, whose shape model is from \cite{willner2014phobos}. (b) Relative $L_2$ error between the FEM solution with the DtN method and the spectral-method solution from \cite{myhill2025} at the outer spherical boundary $r=b$.}
\label{fig:phobos}
\end{figure}
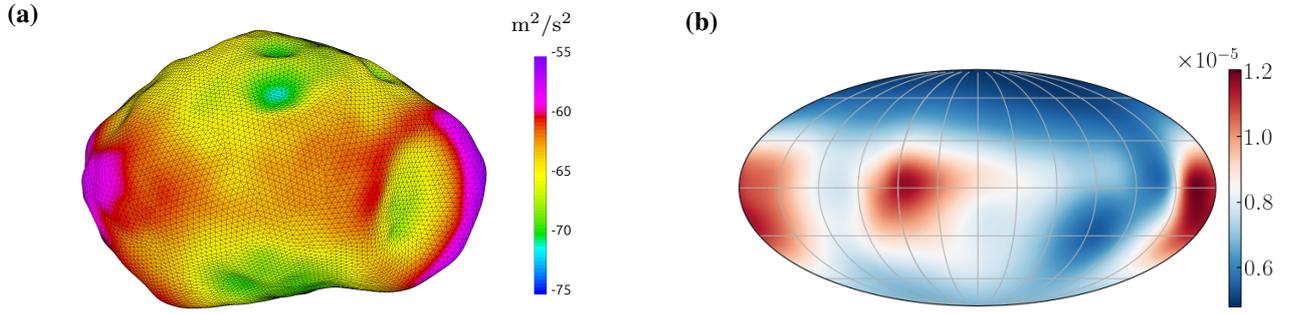

The second application is to the moon Phobos of Mars. We adopt a shape model for Phobos from \citep{willner2014phobos} whose geometry is displayed in Fig.\ref{fig:phobos}(a). 
This geometric shape exhibits significant asymmetry and irregular topographic features. 
We assume a constant density $\rho=1860 \mathrm{kg/m^3}$, a generally-accepted average value. To compute the gravitational potential via FEM, a buffer layer for applying the 
DtN method is chosen with its outer radius being $b=1.5R$, where the length scale $R=11\mathrm{km}$ 
approximates the average radius of Phobos. 

The solution of gravity potential on the surface of Phobos is shown in Fig.\ref{fig:phobos}(a), which is highly asymmetric and higher at regions with larger altitudes. 
To confirm the validity of the FEM algorithm with the DtN method, the solution at the outer spherical boundary $r=b$ is compared in Fig.\ref{fig:phobos}(c) with that determined  by \cite{myhill2025} using a hybrid pseudo-spectral/spectral element method that builds on and refines the earlier method of \cite{maitra2019non}. The relative 
difference between the two solutions on the boundary sphere is of order $10^{-5}$, though we note that this can be reduced further by either refining the finite-element 
mesh or increasing the polynomial order.

\section{Conclusions}\label{sec:conclusions}
Finite-element formulations of self-gravitating problems are fundamentally constrained by how the unbounded exterior is represented. The results presented show that treating the exterior through explicit domain enlargement is possible but sub-optimal. Approaches based on spherical harmonic representations, such as the Dirichlet-to-Neumann and multipole expansion methods, address this mismatch directly by 
accurately approximating the exterior behaviour on a truncated boundary. 

For the benchmark problem considered, both DtN and multipole methods reduce the global error by several orders of magnitude with relatively small spherical harmonic cut-offs, and 
offer significant speedups over the solution on a sufficiently large mesh using simple homogeneous boundary conditions. 
These differences are amplified in realistic and time-dependent applications. In the linearised gravity problem relevant to coupled elasto-gravitational simulations, both methods maintain their accuracy and convergence properties, but the DtN approach benefits from the fact that the boundary operator is independent of the source term and can be assembled once and reused. This leads to a clear efficiency advantage when solves are repeated over many time steps or load configurations. The favourable parallel scaling performance observed for both approaches further shows that the present numerical implementation does not limit scalability. The successful applications of the DtN method to layered Earth models and to the irregular geometry of Phobos demonstrate its robustness beyond idealised configurations and supports its use within large-scale, high-fidelity geophysical simulations. 

All necessary codes have been provided to implement the methods discussed within the MFEM framework, while 
the same approach could be readily adapted within libraries such as deal.ii \citep[][]{bangerth2007deal} which provide the necessary element-level access. 
We have also described a practical approach for the implementation of the DtN and multipole methods within  higher-level finite element 
packages such as FEniCS  and Firedrake by combining their convenient unified form language  with a low-rank and matrix-free representation of 
the associated operators that can be implemented within PETSc or similar linear algebra libraries.

\subsection{Acknowledgments}

\begin{acknowledgments}
ZY and DA have been supported through the Natural Environment
Research Council grant NE/X013804/1. AM acknowledges the receipt of a Gates Cambridge 
scholarship as well as NERC DTP grant. 
\end{acknowledgments}

\subsection{Data Availability}
There is no data associated with this paper. The numerical codes we have developed can be accessed through \url{https://github.com/Zih0770/Self-gravitation_FEM}.

\bibliographystyle{gji}
\bibliography{ref}

\appendix

\section{Solution of Poisson's equation in 2D}\label{app:2d}

The methods discussed within the paper apply also to the solution of Poisson's equation in $\R^{2}$. Within 
this appendix we summarise the necessary modifications, while the corresponding methods have
also been implemented within the codes we have developed and provide. 

\subsection{The DtN method}

As with the 3D problem we let $M$ denote the domain in which the density is non-zero, and $B$ be a larger ball of radius $b$ that contains
$M$. The exterior solution of Poisson's equation in $\R^2$ can be written as a Fourier expansion
\begin{equation}
    \phi(r,\theta)=A_0 + C_0 \ln{r} + \sum_{n=1}^{\infty}\left(\frac{b}{r}\right)^n \left[A_n\cos{(n\theta)}+B_n\sin{(n\theta)}\right],
\end{equation}
with coefficients
\begin{equation}
    A_n = \frac{1}{\pi b}\int_{\partial B}\cos{(n\theta)}\,\phi\dif S,\quad B_n = \frac{1}{\pi b}\int_{\partial B}\sin{(n\theta)}\,\phi\dif S.
\end{equation}
Note that for the 2D problem, the potential need not be bounded as $r\rightarrow\infty$ due to the logarithmic term, but its normal derivative does tend to zero. 
From this expression, we can determine the normal derivative of the potential on $\partial B$ in terms of its Dirichlet data. Putting this 
into the weak form of the problem in eq.(\ref{eq:weak}) we then arrive at
\begin{equation}
   \frac{1}{4\pi G} \int_B\nabla \psi\cdot\nabla \phi + \frac{1}{4\pi G}\sum_{n=1}^{\infty}n\pi\left(A^\dagger_n A_n+B^\dagger_n B_n\right) = -\int_M\psi \rho\dif\mathbf{x} + \frac{1}{2\pi b}\int_M \rho \dif\mathbf{x}\int_{\partial B}\psi\dif S, 
\end{equation}
where the superscript $(\cdot)^\dagger$ denotes the Fourier coefficient for the test function $\psi$ and the value of $C_0$ has been determined by letting the test function 
equal a constant. As with the 3D problem, within the DtN method a symmetric bilinear form is added to the problem that is expressed in terms of the spectral expansion 
of the trial and test functions. Numerically, this spectral expansion is truncated at a finite degree, and a method very similar to that used in 3D can be employed to discretise the form efficiently within the MFEM framework.

\subsection{The multipole method}

The fundamental solution of Poisson equation in 2D takes a different form,
\begin{equation}
    \Phi(\mathbf{x},\mathbf{x'})=\frac{1}{2\pi}\mathrm{ln}|\mathbf{x}-\mathbf{x'}|,
\end{equation}
which admits the Fourier expansion
\begin{equation}\label{eq:fund_2d}
    \Phi(\mathbf{x},\mathbf{x'})=\frac{1}{2\pi}\left[\mathrm{ln}(r)-\sum_{n=1}^{\infty}\frac{1}{n}\left(\frac{r'}{r}\right)^{n}\left(\cos{(n\theta')}\cos{(n\theta)}+\sin{(n\theta')}\sin{(n\theta)}\right)\right].
\end{equation}
Using this result, we can readily obtain solutions of the exterior problem for both the static and linearised Poisson problems 
in terms of a multipole expansion. For the static problem, the necessary weak form is given by
\begin{align}
   \frac{1}{4\pi G} \int_B \nabla \psi \cdot \nabla \phi 
    &= -\int_M \psi \rho\dif\mathbf{x} 
    + \frac{1}{2\pi b}\int_M \rho\dif\mathbf{x}\int_{\partial B}\psi\dif S + \sum_{n=1}^{\infty}\left[
    \int_{\partial B}\cos (n\theta)\,\psi\dif S\,\,\frac{1}{2\pi b}\int_M\left(\frac{r}{b}\right)^n\cos (n\theta)\,\rho\dif\mathbf{x}\right.\notag\\
    &+\left.\int_{\partial B}\sin (n\theta)\,\psi\dif S\,\,\frac{1}{2\pi b}\int_M\left(\frac{r}{b}\right)^n\sin (n\theta)\,\rho\dif\mathbf{x}
    \right], 
\end{align}
while that for the linearised problem is similar. Both methods have been implemented within the MFEM framework following the 
approach discussed for the corresponding 3D problem. 

\end{document}